\documentclass[12pt]{article}
\usepackage{fullpage,graphicx,psfrag,amsmath,amsfonts,verbatim}
\usepackage[small,bf]{caption}
\usepackage{amssymb}
\graphicspath{{figures/}}

\usepackage{listings}
\usepackage{longtable}
\usepackage{subcaption}

\usepackage{color}

\definecolor{codegreen}{rgb}{0,0.6,0}
\definecolor{codegray}{rgb}{0.5,0.5,0.5}
\definecolor{codepurple}{rgb}{0.58,0,0.82}
\definecolor{backcolour}{rgb}{0.95,0.95,0.98}

\lstdefinestyle{mystyle}{
      backgroundcolor=\color{backcolour},
      keywordstyle=\color{magenta},
      numberstyle=\tiny\color{codegray},
      stringstyle=\color{codepurple},
      basicstyle=\ttfamily\small,
      breakatwhitespace=false,
      breaklines=true,
      captionpos=t,
      keepspaces=true,
      numbers=left,
      numbersep=5pt,
      showspaces=false,
      showstringspaces=false,
      showtabs=false,
      tabsize=2,
}

\definecolor{seagreen}{rgb}{0.18, 0.55, 0.34}
\definecolor{mediumviolet-red}{rgb}{0.78, 0.08, 0.52}
\definecolor{khaki}{rgb}{0.94, 0.9, 0.55}

\lstdefinelanguage{mypython}
{
      keywords=[1]{from, import, as, assert, not, print, nonneg, PSD, axis},
      keywordstyle=[1]{\color{mediumviolet-red}},
      keywords=[2]{cp, lo, pl, cvxpy, Variable, Parameter,
      sqrt, exp, numpy, np, Problem, Minimize, Maximize, value, solve, inner,
      sum, multiply, arange, range, norm1, norm2, norm_inf, abs, square,
      diagonal, outer, pos, hstack, power},
      keywordstyle=[2]{\color{seagreen}},
      upquote=true,
      showstringspaces=false,
      basicstyle=\ttfamily,
      columns=fullflexible,
      keepspaces=true,
      emph={True,False,def,return,float,class,match,switch,len},
      emphstyle={\color{seagreen}},
      belowskip=1em,
      aboveskip=1em,
      morecomment=[l]{\#}
}

\newcommand{\BEAS}{\begin{eqnarray*}}
    \newcommand{\EEAS}{\end{eqnarray*}}
    \newcommand{\BEA}{\begin{eqnarray}}
    \newcommand{\EEA}{\end{eqnarray}}
    \newcommand{\BEQ}{\begin{equation}}
    \newcommand{\EEQ}{\end{equation}}
    \newcommand{\BIT}{\begin{itemize}}
    \newcommand{\EIT}{\end{itemize}}
    \newcommand{\BNUM}{\begin{enumerate}}
    \newcommand{\ENUM}{\end{enumerate}}

    \newcommand{\BA}{\begin{array}}
    \newcommand{\EA}{\end{array}}

    \newcommand{\eg}{{\it e.g.}}
    \newcommand{\ie}{{\it i.e.}}

    \newcommand{\ones}{\mathbf 1}

    \newcommand{\reals}{{\mbox{\bf R}}}

    \newcommand{\symm}{{\mbox{\bf S}}}  

    \newcommand{\diag}{\mathop{\bf diag}}

    \newcommand{\Expect}{\mathop{\bf E{}}}

    \newcommand{\var}{\mathop{\bf var}}
    \newcommand{\cov}{\mathop{\bf cov}}

    {\begin{quote}}{\end{quote}}

    \makeatletter
    \long\def\@makecaption#1#2{
       \vskip 9pt
       \begin{small}
       \setbox\@tempboxa\hbox{{\bf #1:} #2}
       \ifdim \wd\@tempboxa > 5.5in
            \begin{center}
            \begin{minipage}[t]{5.5in}
            \addtolength{\baselineskip}{-0.95pt}
            {\bf #1:} #2 \par
            \addtolength{\baselineskip}{0.95pt}
            \end{minipage}
            \end{center}
       \else
        \hbox to\hsize{\hfil\box\@tempboxa\hfil}
       \fi
       \end{small}\par
    }
    \makeatother

    \newcounter{oursection}

    \newcounter{lecture}

\newcommand{\sigmahat}{\hat{\Sigma}}

\usepackage{mathtools}
\usepackage{adjustbox}
\usepackage{booktabs}
\usepackage{csquotes}

\usepackage{hyperref}
\hypersetup{ colorlinks   = true,    
urlcolor     = blue,    
linkcolor    = blue,    
citecolor    = red      
}

\title{Markowitz Portfolio Construction at Seventy}
\author{Stephen Boyd\footnote{Alphabetical order.}
\and Kasper Johansson
\and Ronald Kahn
\and Philipp Schiele
\and Thomas Schmelzer
}

\begin{document}
\maketitle

\begin{abstract}
More than seventy years ago Harry Markowitz formulated
portfolio construction as an optimization problem that trades off
expected return and risk,
defined as the standard deviation of the portfolio returns.
Since then the method has been extended to include many practical 
constraints and objective terms, such as transaction cost or leverage 
limits.  
Despite several criticisms of Markowitz's method, 
for example its sensitivity to poor forecasts of the return statistics, it has become
the dominant quantitative method for portfolio construction in practice.
In this article we describe an extension of Markowitz's method
that addresses many practical effects and gracefully handles the 
uncertainty inherent in return statistics forecasting.  
Like Markowitz's original formulation, the extension is also a convex 
optimization problem, which can be solved with high reliability and 
speed.
\end{abstract}

\clearpage
\tableofcontents
\clearpage

\section{Introduction}

Harry Markowitz's 1952 paper \emph{Portfolio Selection}~\cite{markowitz1952portfolio} 
was a true breakthrough in our understanding of and approach to investing.  
Before Markowitz there was (almost) no
mathematical approach to investing.  As a 25-year-old graduate student,
Markowitz founded modern portfolio theory, and methods inspired by him would
become the most widely used portfolio construction practices over the next 70
years (and counting).

Before Markowitz, diversification and risk were fuzzy concepts.
Investors loosely connected risk to the probability of loss, but with no
analytical rigor around that connection.  Ben Graham, who along with David Dodd
wrote \emph{Security Analysis}~\cite{graham2009security}, 
once commented that investors should own ``a minimum of
ten different issues and a maximum of about thirty''~\cite{graham1973intelligent}.

There were a few precursors, such as an article by de Finetti, that contained
some similar ideas before Markowitz; see \cite{deFinetti1940,rubinstein2006bruno} 
for a discussion
and more of the history of mathematical formulation of portfolio construction.
Another notable precursor is
John Burr Williams' 1938 \emph{Theory of Investment Value}~\cite{williams1938theory}.
He argued that the value of a company was the present value of future
dividends.  His book is full of mathematics, and Williams predicted that
``mathematical analysis is a new tool of great power, whose use promises to lead
to notable advances in investment analysis''.  
That prediction came true with Markowitz's work.
Indeed, Markowitz considered Williams' book as part of his inspiration.
According to Markowitz, ``the basic concepts of portfolio theory came to me one
afternoon in the library while reading John Burr Williams' \emph{Theory 
of Investment Value}''.

For many years, the lack of data and accessible 
computational power \cite{markowitz2019aqr}
rendered Markowitz's ideas impractical, despite his pragmatic approach.  In
1963, William Sharpe published his market model~\cite{sharpe1963}, designed to speed up
the Markowitz calculations. This model was a 
one-factor risk model (the factor was the market return), 
with the assumption that all residual
returns are uncorrelated. His paper stated that solving a 100-asset problem on
an IBM 7090 computer required 33 minutes, but his simplified risk model reduced
it to 30 seconds. He also commented that computers could only handle 249
assets at most with a full covariance matrix, but 2000 assets with the
simplified risk model.  Today such a problem can be solved in microseconds; we
can routinely solve problems with tens of thousands of assets and substantially
more factors in well under one second.

Markowitz portfolio construction has thrived for 
many years in spite of claims of
various alleged deficiencies.  These have included the method's sensitivity to
data errors and estimation uncertainty, its single-period nature to handle what is
fundamentally a multi-period problem, its symmetric definition of risk, and
its neglect of higher moments like skewness and kurtosis.  We will address these
alleged criticisms and show that standard techniques in modern approaches to
optimization effectively deal with them without altering Markowitz's vision for
portfolio selection.  

In 1990 Markowitz was awarded the Nobel Memorial Prize in Economics
for his work on portfolio theory, shared with Merton Miller and William Sharpe.
For more light on the fascinating historic details we recommend
an interview with Markowitz~\cite{markowitz2019aqr},
his acceptance speech for the Nobel Prize~\cite{markowitz1991},
and his remarks in the introduction to the \emph{Handbook of Portfolio
Construction}~\cite{GuerardHandbook}.

\subsection{The original Markowitz idea}
Markowitz identified two steps in the portfolio selection process. In a first step,
the investor forms beliefs about the expected returns of the assets,
expressed as a vector $\mu$, and their covariances, expressed
as a covariance matrix $\Sigma$, which gives
the volatilities of asset returns and the correlations among them.
These beliefs are the core inputs for the second step, which is the optimization
of the portfolio based on these quantities.

He introduced the expected returns--variance of returns (E--V) rule,
which states that an investor desires to achieve the maximum expected
return for a portfolio while keeping its variance or risk below a given threshold.
Convex programming was not a well developed field at that time,
and Markowitz used a geometric interpretation in the space of
portfolio weights~\cite{markowitz1952portfolio} to solve
the problem we would now express as
\BEQ \label{e-markowitz-original}
\begin{array}{ll}
\mbox{maximize}   &  \mu^T w\\
\mbox{subject to} &  w^T \Sigma w \leq (\sigma^\text{tar})^2,\\
                 &  \ones^T w = 1,\\
\end{array}
\EEQ
with variable $w\in\reals^n$, the set of portfolio weights,
where $\ones$ is the vector with all entries one.
The data in the problem are $\mu\in\reals^n$,
the vector of expected asset returns, and
$\Sigma$, the $n\times n$ covariance matrix of asset returns.
The positive parameter $\sigma^\text{tar}$ is the target portfolio return
standard deviation or volatility.
(We define the weights and describe the problem 
more carefully in \S\ref{s-portfolio-and-holdings}.)

There are many other ways to formulate the trade-off of expected return and 
risk as an optimization problem \cite{boydconvex,aps2023mosek}.
One very popular method maximizes the \emph{risk-adjusted return},
which is the expected portfolio return minus its variance, scaled by
a positive \emph{risk-aversion parameter}.
This leads to the optimization problem~\cite{grinold2000active}
\BEQ \label{e-markowitz-risk-adj}
\begin{array}{ll}
\mbox{maximize}   &  \mu^T w - \gamma w^T \Sigma w\\
\mbox{subject to} &  \ones^T w = 1,\\
\end{array}
\EEQ
where $\gamma$ is the risk-aversion parameter that controls
the trade-off between risk and return.
Both problems \eqref{e-markowitz-original} and \eqref{e-markowitz-risk-adj} 
give the full trade-off curve of Pareto optimal
weights, as $\sigma^\text{tar}$ or $\gamma$ vary from $0$ to $\infty$
(although \eqref{e-markowitz-original} can be infeasible when 
$\sigma^\text{tar}$ is too small).
One advantage of the first formulation \eqref{e-markowitz-original}
is that the parameter $\sigma^\text{tar}$ that controls the volatility 
is interpretable as, simply, the target risk level.
The risk-aversion parameter $\gamma$ appearing in \eqref{e-markowitz-risk-adj}
is less interpretable.
We will have more to say about the parameters that control 
trade-offs in portfolio construction in \S\ref{s-cvx-formulation}.

Both problems \eqref{e-markowitz-original}  and
\eqref{e-markowitz-risk-adj} have analytical solutions.
For example the solution of \eqref{e-markowitz-risk-adj} is given by
\[
w^{\star} = \frac{1}{2\gamma} \Sigma^{-1} (\mu+\nu^{\star}\ones), \qquad
\nu^{\star} = \frac{2\gamma - \ones^T \Sigma^{-1}\mu}{\ones^T \Sigma^{-1} \ones}.
\]
(The scalar $\nu^\star$ is the optimal dual variable~\cite[Chap.~5]{boydconvex}.)
We note here the appearance of the inverse covariance matrix.  
To compute $w^\star$ we would not compute the inverse, but rather solve 
two sets of equations to find 
$\Sigma^{-1}\mu$ and $\Sigma^{-1}\ones$~\cite{boyd2018introduction}.
Still, the appearance of $\Sigma^{-1}$ in the expressions for the 
solutions give us a hint that the method can be sensitive 
to the input data when the covariance matrix $\Sigma$ is nearly singular.
These analytical formulas can also be used to back out so-called 
implied returns, \ie, the mean $\mu$ for which a given portfolio is 
optimal.  For example the \emph{market implied return} $\mu^\text{mkt}$ is
the return for which the optimal weights are the market weights, \ie, proportional
to asset capitalization.

Both formulations \eqref{e-markowitz-original}  and
\eqref{e-markowitz-risk-adj} are referred to as the basic Markowitz
problem, or \emph{mean-variance optimization}, since they both trade off 
the mean and variance of the portfolio return.
In his original paper Markowitz also noted that additional constraints can 
be added to the problem, specifically the constraint that $w\geq 0$
(elementwise), which means the portfolio is long-only, \ie, it does not 
contain any short positions.
With this added constraint, the two problems above do not have simple 
analytical solutions.  But the formulation \eqref{e-markowitz-risk-adj},
with the additional constraint $w \geq 0$,
is a quadratic program (QP), a type of convex optimization problem 
for which numerical
solvers were developed already in the late 1950s~\cite{wolfe1959quadratic}.
In that early paper on QP, solving the Markowitz 
problem \eqref{e-markowitz-risk-adj} with the long-only constraint $w \geq 0$
was listed as a prime application.
Today we can solve either formulation reliably, with essentially 
any set of convex portfolio constraints.

Since the 1950s we have seen a truly stunning increase in computer power,
as well as the development of convex optimization methods that are 
fast and reliable, and high-level languages that allow users to 
express complex convex optimization problems in a few lines of clear code.
These advances allow us to extend Markowitz's formulation to 
include a large number of practical constraints and additional
terms, such as transaction cost or the cost incurred when holding short
positions.  In addition to directly handling a number of practical issues,
these generalizations of the basic Markowitz method also address
the issue of sensitivity to the input data $\mu$ and $\Sigma$.
This paper describes one such generalization of the basic Markowitz problem,
that works well in practice.

Out of respect for Markowitz, and because the more generalized formulation
we present here is nothing more than an extension of his original idea, 
we will refer to these more complex portfolio
construction methods also as Markowitz methods.
When we need to distinguish the extension of Markowitz's portfolio construction
that we recommend from the basic Markowitz method, we refer to it as 
Markowitz++.  (In computer science, the post-script ++ denotes 
the successor.)

\subsection{Alleged deficiencies}\label{s-alleged-deficiencies}
The frequent criticism of Markowitz's work is a testament to its importance.
These criticisms usually fall into one or more of the following (related) categories.

\paragraph{It's sensitive to data errors and estimation uncertainty.}
The sensitivity of Markowitz portfolio construction to input data is
well documented~\cite{muller1993,michaud2008efficient,schmelzer2013seven,
brandt2010portfolio, chen2016efficient},
and already hinted at 
by the inverse covariance that appears in the analytical solutions of
the basic Markowitz method.
This sensitivity, coupled with the challenge of estimating 
the mean and covariance of the return, leads to
portfolios that exacerbate errors or deficiencies in the input data
to find unrealistic and poorly performing portfolios.
Some authors argue that choosing a portfolio by optimization,
as Markowitz's method does,
is essentially an estimation-error maximization method.  
This is still a research topic that draws much attention. In
the recent papers \cite{goldberg2022dispersion, shkolnik2023markowitz} 
the authors quantify how the (basic) Markowitz portfolio is affected by 
estimation errors in the covariance matrix.

This criticism is justified, on the surface. Markowitz portfolio construction can 
perform poorly when it is na\"{i}vely implemented, for example by using
empirical estimates of mean and covariance on a trailing window of past returns.
But the critical practical issues of taming sensitivity and 
gracefully handling estimation errors are readily addressed using 
techniques such as regularization and robust optimization, described in
more detail in \S\ref{s-robust-opt}.

\paragraph{It implicitly assumes risk symmetry.}
Markowitz portfolio construction uses variance of the portfolio return as 
its risk measure.  With this risk measure a portfolio return well above the mean is 
just as bad as one that is well below the mean, whereas the 
former is clearly a good event, not a bad one.  This observation should at least make
one suspicious of the formulation, and has motivated a host of 
proposed alternatives, such as defining the risk taking into account only
the downside~\cite[Chap.~IX]{markowitz1959portfolio}.
This criticism is also valid, on the surface.  But when the parameters are 
chosen appropriately, and the data are reasonable, 
portfolios constructed from mean-variance optimization do not suffer from
this alleged deficiency.

\paragraph{We should maximize expected utility.}
A more academic version of the previous criticism is that portfolios 
should be constructed by maximizing the expected value of 
a concave increasing utility function of the portfolio return
\cite{von1947theory}. 
The utility in mean-variance optimization (with risk-adjusted return objective)
is $U(R) = R-\gamma R^2$, where $R$ is the portfolio return.
This utility function is concave, but only increasing for $R<1/(2\gamma)$;
above that value of return, it decreases, putting us in the awkward 
position of seeming to prefer smaller returns over larger ones.

This criticism is also valid, taken at face value; the quadratic utility above 
is indeed not increasing.  Markowitz himself addressed the issue in a 1979 paper
with H. Levy that argued that while mean-variance optimization does not appear to 
be the same as maximizing an expected utility, it is a very good approximation;
see~\cite{levy1979approximating} and \cite[Chap.~2]{markowitz2014risk}.
But in fact it turns out that Markowitz portfolio construction
\emph{does} maximize the expected value of a concave increasing utility function.
Specifically if we model the returns as Gaussian, and use the exponential
utility $U(R)=1-\exp(-\gamma R)$, then the expected utility is
the risk-adjusted return, up to an additive constant~\cite{luxenberg2023portfolio}.
In other words, Markowitz portfolio construction \emph{does} maximize expected utility
of portfolio return, for a specific concave increasing utility function and 
a specific asset return distribution.

\paragraph{It considers only the first and second moments of the return.} 
Mean-variance optimization naturally only
considers the first two moments of the distribution. It would seem that taking
higher moments like skewness and kurtosis into account might better describe 
investor preferences \cite{Cajas2022, Palomar2021}.
This, coupled with the fact that the tails of 
asset returns are not well modeled by a Gaussian distribution~\cite{fama1965behavior},
suggests that portfolio construction should consider higher moments 
than the first and second.

While it is possible to construct small academic examples where mean-variance 
optimization does poorly due to its neglect of higher moments, simple mean-variance
optimization does very well on practical problems.
In~\cite{luxenberg2023portfolio} the authors extend Markowitz by maximizing exponential 
utility, but with a more complex Gaussian mixture model of asset returns.  Such a 
distribution is general, in that it can approximate any distribution.  Their method
evidently handles higher moments, but empirically gives no boost in 
performance on practical problems.

Markowitz himself addressed the common misconception that he labeled the
``Great Confusion''~\cite{markowitz2019aqr, markowitz2014risk, Markowitz99, markowitz2009harry}, 
stating that Gaussian returns are merely a sufficient but not a necessary condition
on the return distribution for mean-variance optimization to work well and
that mean and variance are good approximations for expected utility.

\paragraph{It's a greedy method.}
Portfolios are generally not just set up and then held for one investment period;
they are rebalanced, and sometimes often.
Problems in which a sequence of decisions are made,
based on newly available information, are more accurately modeled not as simple
optimization problems, but instead as stochastic control problems,
also known as sequential decision making under uncertainty 
\cite{kochenderfer2015decision, kochenderfer2022algorithms,
bellman1966dynamic,bertsekas2012dynamic}.
In the context of stochastic control, methods that take into 
account only the current decision and not future ones 
are called \emph{greedy}, and in some cases can perform very poorly.
This criticism is also, on its face, valid.  Using Markowitz portfolio
construction repeatedly, as is always done in practice, is a greedy method.

We can readily counter this criticism.  First, in the special case with
risk-adjusted return and quadratic transaction costs, and no additional
constraints, the stochastic optimal policy can be worked out, and 
coincides with a single-period Markowitz 
portfolio \cite{grinold2020advances,barratt2021stochastic}.
This suggests that when other constraints are present, and the 
transaction cost is not quadratic, the (greedy) Markowitz method should not
be too far from stochastic optimal.

Second, there are extensions of 
Markowitz portfolio construction, called
\emph{multi-period} methods, that plan a sequence of trades over a horizon,
and then execute only the first trade;
see, \eg, \cite{BoydKahnMultiPeriod, li2022multi}. 
These multi-period methods can work better than so-called single-period methods,
for example when a portfolio is transitioning between two managers, or being set up
or liquidated over multiple periods.  
But in almost all other cases, single-period methods work just as well
as multi-period ones. 

The third response to this criticism more directly addresses the question. In
the paper \emph{Performance Bounds and Suboptimal Policies for Multi-Period
Investment} ~\cite{boyd2013performance}, the authors develop bounds on how well
a full stochastic control trading policy can do, and show empirically that
single-period Markowitz trading essentially does as well as a full stochastic
control policy (which is impractical if there are more than a handful of
assets).  So while there are applications where greedy policies do much more
poorly than a true stochastic control policy, it seems that multi-period trading
is not one of them.

\subsection{Robust optimization and regularization}\label{s-robust-opt}
Here we directly address the question of sensitivity of Markowitz portfolio
construction to the input data $\mu$ and $\Sigma$.
As mentioned above, the basic methods are indeed sensitive to these parameters.
But this sensitivity can be mitigated and tamed using techniques 
that are widely used in other applications and fields,
robust optimization and regularization.

\paragraph{Robust optimization.}
Modifying an optimization-based method to make it more robust to data uncertainty 
is done in many fields, using techniques that have differing names.
When optimization is used in almost any application, some of the data
are not known exactly, and solving the optimization problem
without recognizing this uncertainty, for example by using some kind of
mean or typical values of the parameters, can lead to very poor practical
performance.
\emph{Robust optimization} is a subfield of optimization that develops methods
to handle or mitigate the adverse effects of
parameter uncertainty; see, \eg, \cite{ben2009robust, tutuncu2004robust, gabrel2014recent, ben2002robust, bertsimas2011theory, lobo2000robust}.
These methods tend to fall in one of two approaches: statistical or worst-case 
deterministic.
In a statistical model, the uncertain parameters are modeled as
random variables and the goal is to optimize the expected value
of the objective under this distribution, leading to a 
\emph{stochastic optimization problem}~\cite{shapiro2021lectures}, 
\cite[Chap.~6.4.1]{boydconvex}.  A worst-case deterministic
uncertainty model posits a set of possible values for parameters,
and the goal is to optimize the worst-case value of the objective
over the possible parameter values~\cite{beyer2007robust}, \cite[Chap.~6.4.2]{boydconvex}. 
Another name for worst-case robust optimization is \emph{adversarial optimization},
since we can model the problem as us choosing values for the variables to
obtain the best objective,
after which an adversary chooses the values of the parameters so as to achieve
the worst possible objective.
Worst-case robust optimization has many variations and goes by many names.  
For example when the set of possible parameter values is finite, they are called 
\emph{scenarios} or \emph{regimes}, and optimizing for the worst-case scenario
is called \emph{worst-case scenario optimization}.
While these general approaches sound quite different, they often lead
to very similar solutions, and both can work well in applications.
Robust optimization methods work by modifying the objective or 
constraints to model the possible variation in the data.

One very successful application of robust optimization is in
\emph{robust control}, where a control system is designed so that
the control performance is not too sensitive to changes in the system dynamics
\cite{zhou1998essentials,khalil1996robust}.  So-called linear quadratic 
optimal control was developed around 1960, and used in many applications.
Its occasional sensitivity to the data (in this case, the 
dynamic model of the system being controlled) was noted then; by the early 1990s
robust control methods were developed, and are now very widely used.

\paragraph{Regularization.}  Regularization is another term for 
methods that modify an optimization problem to mitigate sensitivity to data.
It is almost universally used in statistics and machine learning 
when fitting models to data.
Here we fit the parameters of a model to some given training data, 
accounting for the fact that the training data set could have been 
different~\cite{tikhonov1977solutions,hastie2009elements}. 
This process of regularization can be done explicitly by adding a 
penalty term to the objective, and also implicitly by adding constraints
to the problem that prevent extreme outcomes.  
Regularization can often
be interpreted as a form of robust optimization; see, \eg, 
\cite[Chap.~6.3--6.4]{boydconvex}.

\paragraph{The high level story.}
Robust optimization and regularization both follow the same high level story,
and both can be applied to the Markowitz problem.
The story starts with a basic optimization-based method that relies on data 
that are
not known precisely.  We then modify the optimization problem, often by adding 
additional objective terms or constraints.  Doing this \emph{worsens} the 
in-sample performance.  But if done well, it \emph{improves} out-of-sample
performance.  Roughly speaking, robustification and regularization tell
the optimizer to not fully trust the data, and this serves it well out-of-sample.

In portfolio construction a long-only constraint can be interpreted as a
form of regularization~\cite{jagannathan2003longonly}.
A less extreme version is to impose a leverage limit, which can help avoid
many of the data sensitivity issues.
We will describe below some effective and simple robustification methods for 
portfolio construction.

Regularization can (and should) also be applied to the forecasting of the mean and 
covariance in Markowitz portfolio construction.  
The Black-Litterman approach to estimating the mean returns
regularizes the estimate toward the market implied return \cite{black1990asset}.
A return covariance estimate can be regularized using \emph{shrinkage},
another term for regularized estimation in statistics \cite{Ledoit2004}.

\subsection{Convex optimization}
Over the same 70-year period since Markowitz's original work, 
there has been a parallel advance in
mathematical optimization, and especially convex optimization, not to mention
stunning increases in available computer power.
Roughly speaking, convex optimization problems are mathematical 
optimization problems that satisfy certain mathematical properties.
They can be solved reliably and efficiently, even when they involve a very large
number of variables and constraints, and involve nonlinear, even 
nondifferentiable, functions \cite{boydconvex}.

Shortly before Markowitz published his paper on portfolio selection,
George Dantzig developed the simplex method~\cite{dantzig1951maximization},
which allowed for the efficient solution of linear programs.
In 1959, Wolfe~\cite{wolfe1959quadratic} extended
the simplex method to QP problems, citing Markowitz's work as
a motivating application.
This close connection between portfolio construction and optimization 
was no coincidence, since Dantzig and Markowitz were colleagues at RAND.

Since then, the field of convex optimization has grown tremendously.
Today, convex optimization is a mature field with a large body of theory,
algorithms, software, and applications~\cite{boydconvex}. 
Being able to solve optimization problems reliably and efficiently
is crucial for portfolio construction, especially for back-testing or simulating
a proposed method on historical or synthesized data, where portfolio 
construction has to be carried many times.
Thus, any extension of the Markowitz objective or additional constraints
should be convex to ensure tractability.
As we will see, this is hardly a limitation in practice.

\paragraph{Solvers.}
The dominant convex optimization problem form is now the \emph{cone program}, 
a generalization of linear programming that handles nonlinear objective terms 
and constraints \cite{NesterovNemirovskyConic,boydconvex,boydSOCP,boydSDP}.
There are now a number of reliable and efficient solvers for such problems,
including open-source ones
like ECOS~\cite{domahidi2013ecos}, Clarabel~\cite{clarabeldocs}, and SCS~\cite{o2016conic},
and commercial solvers 
such as MOSEK~\cite{aps2020mosek}, GUROBI~\cite{gurobi}, and CPLEX~\cite{cplex2009v12}. 
A recent open-source solver for QPs is OSQP~\cite{stellato2020osqp}.

\paragraph{Domain-specific languages.}
Convex optimization is also now very accessible to
practitioners, even those without a strong background in the mathematics or 
algorithms of convex optimization, thanks to
high-level domain-specific languages (DSLs) for convex optimization, such
as CVXPY~\cite{agrawal2018rewriting, cvxpy}, CVX~\cite{grant2014cvx},
Convex.jl~\cite{udell2014convex}, CVXR~\cite{fu2017cvxr}, and YALMIP~\cite{lofberg2004yalmip}. 
These DSLs make it easy to specify complex, but convex, optimization problems
in a natural, human readable way.  The DSLs transform the 
problem from the human readable form to a lower level form (often a cone program)
suitable for a solver.
These DSLs make it easy to develop convex optimization based methods,
as well as to modify, update, and maintain existing ones.  As a result,
CVXPY is used at many quantitative hedge funds today,
as well as in many other applications and industries.
The proposed extension of Markowitz's portfolio construction method
that we describe below is a good example of the use of CVXPY.  It is 
a complex problem involving nonlinear and nondifferentiable functions,
but its specification in CVXPY takes only a few tens of lines of clear readable code,
given in appendix~\ref{s-CVXPY}.
The overhead of translating the human readable problem specification
into a cone program is typically small.
Additionally, in some DSLs, such as CVXPY, problems can be 
parametrized~\cite{cvxpylayers2019}, such that they can be solved for a range 
of values of the parameters, making the translation overhead negligible.
Related to DSLs are modeling layers provided by some solver, such as 
MOSEK's Fusion API~\cite{aps2020mosek}, which provides a high-level interface to the
solver. 
Less focused on convex optimization, there are other modeling languages such as
JuMP~\cite{Lubin2023} and Pyomo~\cite{hart2011pyomo,bynum2021pyomo} that do not
verify convexity, but provide flexibility in modeling a wide range of 
optimization problems, including nonconvex ones.

\paragraph{Code generators.}
Code generators
like CVXGEN~\cite{mattingley2012cvxgen} and CVXPYgen~\cite{schallerCVXPYgen}
are similar to DSLs.  They support high level specification of a problem (family)
but instead of directly solving the problem, they generate 
custom low level code (typically C) for the problem that is specified.  
This code can be compiled to
a very fast and totally reliable solver, suitable for embedded real-time applications.
For example, CVXGEN-generated code guides all of SpaceX's Falcon 9 and Falcon Heavy 
first stages to their landings~\cite{blackmore2016autonomous}.

\subsection{Previous work}
The literature on portfolio construction is vast, and focusing on the practical
implementation of Markowitz's ideas, we do not attempt to survey it here in 
detail.
Instead, we highlight only a few major developments that are relevant to our
work. For a detailed overview see,
\eg,~\cite[Chap.~14]{grinold2000active},~\cite[Chap.~6]{narang2009inside}, and
\cite{cornuejols2006optimization, kolm201460}.

Building on Markowitz's framework, the field of portfolio construction has 
undergone substantial evolution. Notable contributions include Sharpe's 
Capital Asset Pricing Model~\cite{sharpe1963} and the Black-Litterman 
model~\cite{black1990asset}. 
A pivotal figure in bringing the field to the forefront of the industry was
Barr Rosenberg, whose research evolved to become the Barra risk
model~\cite{Rosenberg1984, sheikh1996barra}, first used for risk modeling and 
then in portfolio optimization. 
The introduction of risk parity  models~\cite{Maillard2010} brought a focus on 
risk distribution. Additionally, hierarchical risk parity, a recent advancement, 
offers a more intricate approach to risk allocation, considering the 
hierarchical structure of asset correlations~\cite{de2016building}. 
These developments reflect the field's dynamic adaptation to evolving market 
conditions and analytical techniques.

\paragraph{Software.} Dedicated software helped practitioners access the solvers
and DSLs mentioned earlier, and has facilitated the wide acceptance of Markowitz
portfolio construction. A wealth of software packages have been developed for
portfolio optimization, many (if not most) with Python interfaces, both open-source and
commercial. Examples range from simple web-based visualization tools to complex
trading platforms. Here we mention only a few of these software implementations.

On the simpler end Portfolio Visualizer~\cite{portfoliovisualizer} is a
web-based tool that allows users to back-test and visualize various portfolio
strategies. PyPortfolioOpt~\cite{martin2021pyportfolioopt} and Cvxportfolio~\cite{BoydKahnMultiPeriod} are Python packages
offering
various portfolio optimization techniques. PyPortfolioOpt includes mean-variance optimization,
Black-Litterman allocation~\cite{black1990asset}, and more recent alternatives like the Hierarchical Risk
Parity algorithm~\cite{de2016building}, while
Cvxportfolio~\cite{BoydKahnMultiPeriod} supports multi-period strategies.
The skfolio~\cite{skfolio} package offers similar functionality to
PyPortfolioOpt, with a focus on interoperability with the
scikit-learn~\cite{scikit-learn} machine learning library.
Another Python implementation is proposed in~\cite{sarmas2020multicriteria}, where the authors
introduce an approach to multicriteria portfolio optimization.
Quantlib~\cite{quantlib} is an alternate open-source software package for modeling,
trading, and risk management. 

The list of commercial software is also extensive. MATLAB's Financial
Toolbox~\cite{brandimarte2013numerical, matlabfintoolbox} includes functions for
mathematical modeling and statistical analysis of financial data, including 
portfolio optimization. Another example is Axioma, which on top of its popular
risk model offers a portfolio optimizer~\cite{axioma}. 

Other software packages include
Portfolio123~\cite{portfolio123},
PortfoliosLab~\cite{portfolioslabwebsite}, and
PortfolioLab by Hudson \& 
Thames~\cite{hudsonthameswebsite}.
Additionally, many solvers, such as MOSEK~\cite{aps2020mosek,aps2023mosek}, 
provide extensive examples of portfolio optimization problems,
making them easy to use for portfolio optimization.

\subsection{This paper}

Our goal is to describe an extension of the basic Markowitz portfolio construction
method that includes a number of additional objective terms and constraints 
that reflect practical issues and address the issue of sensitivity to 
inevitable forecasting errors.  We give a minimal formulation that is both
simple and practical; we make no attempt to list all possible extensions
that a portfolio manager (PM) might wish to add.

While the resulting optimization problem might appear complex,
containing nonlinear nondifferentiable functions, it is convex, which means 
it can be solved reliably and efficiently.  It can also be specified in a DSL
such as CVXPY in just a few tens of lines of clear simple code.
We can solve even large instances of the optimization problem very quickly, 
making it practical to carry out extensive back-testing to predict performance or 
adjust parameter values.
One additional advantage of our formulation is that parameters that need to
be specified are generally more interpretable than those appearing in basic formulations.  
For example a PM specifies a target risk and a target turnover instead of 
some parameters that are less directly related to them.

Most of the material in this paper is not new but scattered across many sources,
in different formats, and indeed in different application fields.
Some of our recommendations are widely accepted and industry standard,
but others are rarely discussed in the literature
and even less commonly used in practice.

The authors bring a diverse set of backgrounds to this paper.  Some of us have
applied Markowitz portfolio construction day-to-day in research, writing, and real
portfolios.  Others approach Markowitz's method from the perspective of
optimization and control in engineering.  
Control systems engineering has a long history and is widely applied 
in essentially all engineering applications.  Most applications of control
engineering use methods based on models that are either wrong or heavily simplified.
While na\"{i}ve implementations of these methods do not work well (or worse),
simple sensible modifications, similar to the ones we describe later 
in this paper, work very well in practice.

These different backgrounds together can
provide a new perspective and bring modern tools to the endeavor Markowitz began.
These techniques have made Markowitz's method even more applicable and useful to
investors.

\paragraph{Software.}
We have created two companion
software packages. One is designed for pedagogical purposes, uses limited
parameter testing and checking, and very closely follows the terminology and notation of
the paper. It is available at 
\begin{center}
    \url{https://github.com/cvxgrp/markowitz-reference}.
\end{center}
The second package is a robust and flexible implementation, which is better
suited for practical use. It is available at  
\begin{center}
    \url{https://github.com/cvxgrp/cvxmarkowitz}.
\end{center}

\paragraph{Outline.} 
In \S\ref{s-portfolio-and-holdings} we set up our notation, define
weights and trades, and describe various objective terms and constraints.
Return and risk forecasts are covered in \S\ref{s-return-and-risk}.
In \S\ref{s-cvx-formulation} we pull together the material of the 
previous two sections to define the (generalized) Markowitz trading problem,
which we refer to as Markowitz++.
In \S\ref{s-experiments} we present some simple numerical experiments 
that illustrate how the extra terms robustify the basic Markowitz trading policy,
and how parameters are tuned via back-testing to improve good performance.

\section{Portfolio holdings and trades} \label{s-portfolio-and-holdings}

This section introduces the notation and terminology for portfolio
holdings, weights, and trades, fundamental objects
in portfolio construction independent of the trading strategy.
We follow the notation of \cite{BoydKahnMultiPeriod}, with the exceptions
of handling the cash weight separately and dropping the time period subscript.

\subsection{Portfolio weights}

\paragraph{Universe.} We consider a portfolio consisting of investments
(possibly short) in $n$ assets, plus a cash account.
We refer to the set of assets we might hold as the \emph{universe} of
assets, and $n$ as the size of the universe.
These assets are assumed to be reasonably liquid, and could include, for example,
stocks, bonds, or currencies.

\paragraph{Asset and cash weights.}
To describe the portfolio investments, we work with the \emph{weights}
or fractions of the total portfolio value for each asset,
with negative values indicating short positions.
We denote the weights for the assets as $w_i$, $i=1, \ldots, n$,
and collect them into a portfolio weight vector
$w =(w_1, \ldots, w_n)\in \reals^n$.
The weights are readily interpreted: $w_i=0.05$ means that $5\%$ of the total
portfolio value is held in asset $i$, and $w_k=-0.01$ means that we hold a short
position in asset $k$, with value $1\%$ of the total portfolio value.
The dollar value of asset $i$ held is $Vw_i$, where $V$ is the total portfolio
value, assumed to be positive.

We denote the weight for the cash account, \ie, our cash value divided
by the portfolio value, as $c$.
If $c$ is negative, it represents a loan.
When $c>0$ we say the portfolio is \emph{diluted} with cash;
when $c<0$, the portfolio is \emph{margined}.
The dollar value of the cash account is $Vc$.

By definition the weights sum to one, so we have
\BEQ\label{e-weight-sum}
\ones^T w + c = 1,
\EEQ
where $\ones$ is the vector with all entries one.
The first term, $\ones^T w$, is the total weight on the non-cash assets,
and we refer to it as the total asset weight.
The cash weight is one minus the total asset weight,
\ie, $c=1-\ones^T w$.

Several portfolio types can be expressed in terms of the holdings.
A \emph{long-only} portfolio is one with all asset weights nonnegative,
\ie, $w\geq 0$ (elementwise).  A portfolio with $c=0$, \ie, no cash holdings,
is called \emph{fully invested}.
In such a portfolio we have $\ones^T w =1$, \ie, the total asset weight is one.
As another example, a \emph{cash-neutral} portfolio is one with $c=1$.
For a cash-neutral portfolio we have $\ones^T w = 0$, \ie,
the total (net) asset weight is zero.

\paragraph{Leverage.}
The \emph{leverage} of the portfolio, denoted $L$, is
\[
L = \sum_{i=1}^n |w_i| = \|w\|_1.
\]
(Several other closely related definitions are also used. Our definition is
commonly referred to as the \emph{gross leverage} \cite{ANG2011102}.) The
leverage does not include the cash account.

In a long-only portfolio, the leverage is equal to the total asset weight.
The \emph{130-30 portfolio}~\cite{leibowitz2009modern}
refers to a fully invested portfolio with leverage $L=1.6$.
For such a portfolio, the total weight of the short positions
(\ie, negative $w_i$) is $-0.3$ and the total weight of the
long positions (\ie, positive $w_i$) is $1.3$.

\paragraph{Benchmark and active weights.}
In some cases our focus is on portfolio performance
relative to a \emph{benchmark portfolio}.
We let $w^\mathrm{b}\in \reals^n$ denote the weights of the benchmark.
Typically the benchmark does not include any cash weight,
so $\ones^T w^\mathrm{b} = 1$.
We refer to $w-w^\mathrm{b}$ as the \emph{active weights} of our portfolio.
A positive active weight on asset $i$, \ie, $w_i-w^\mathrm{b}_i>0$,
means our portfolio is \emph{over-weight} (relative to the
benchmark) on asset $i$;
a negative active weight, $w_i-w^\mathrm{b}_i<0$, means our portfolio
is \emph{under-weight} on asset $i$.

\subsection{Holding constraints and costs}

Several constraints and costs are associated with the portfolio holdings
$w$ and $c$.

\paragraph{Weight limits.}
Asset and cash \emph{weight limits} have the form
\[
w^\text{min}_i \leq w_i \leq w^\text{max}_i, \quad i=1,\ldots, n,
\quad
c^\text{min} \leq c \leq c^\text{max},
\]
where $w^\text{min}$ and $w^\text{max}$ are given vectors of lower
and upper limits on asset weights,
and $c^\text{min}$ and $c^\text{max}$ are given lower and upper limits on
the cash weight.
We write the asset weight inequalities in vector form as $w^\text{min} \leq
w \leq w^\text{max}$.
We have already encountered a simple example:
a long-only portfolio has $w^\text{min}=0$.

Portfolio weight limits can reflect hard requirements,
for example that a portfolio must (by legal or regulatory requirements)
be long-only.
Portfolio weight limits can also be used to avoid excessive
concentration of a portfolio, or limit short positions.
For example, $w^\text{max} = 0.15$ means that our portfolio cannot
hold more than $15\%$ of the total portfolio value in any one asset.
(Here we adopt the convention that in a vector-scalar inequality, the
scalar is implicitly multiplied by $\ones$.)
As another example, $w^\text{min} = -0.05$ means
that the short position in any asset can never exceed $5\%$ of the total
portfolio value.
For large portfolios it is reasonable to also limit holdings
relative to the asset capitalization, \eg, to require that our
portfolio holdings of each asset are no more than $10\%$ of the asset
capitalization.

Weight limits can also be used to capture the portfolio manager's views
on how the market will evolve.  For example, she might insist
on a long position for some assets, and a short position for some others.

When a benchmark is used, we can impose limits on active weights.
For example $|w-w^\text{b}| \leq 0.10$ means that no asset in the portfolio
can be more than $10\%$ over-weight or under-weight.

\paragraph{Leverage limit.}
In addition to weight limits, we can impose a leverage limit,
\BEQ\label{e-leverage-limit}
L \leq L^\text{tar},
\EEQ
where $L^\text{tar}$ is a specified maximum or target leverage value.
(Other authors have suggested including leverage as a penalty term in
the objective, to model leverage aversion and identify the optimal
amount of leverage in the presence of 
leverage aversion \cite{jacobs2013}.)

\paragraph{Holding costs.}
In general a fee is paid to borrow an asset in order to enter a short
position.  Analogously we pay a borrow cost fee for a negative cash weight.
We will assume these holding costs are a linear function of the
negative weights, \ie, of the form
\BEQ\label{e-h-cost}
\phi^\text{hold}(w,c) = (\kappa^\text{short})^T (-w)_+ +
\kappa^\text{borrow} (-c)_+,
\EEQ
where $(a)_+ = \max\{a,0\}$ denotes the nonnegative part,
applied elementwise and in its first use above.
Here $\kappa^\text{short} \geq 0$ is the vector of borrow cost (rates) for
the assets, and $\kappa^\text{borrow}\geq 0$ is the borrow cost for cash.

\paragraph{Other holding constraints.}
There are many other constraints on weights that might be imposed,
some convex, and others not.
A \emph{concentration limit} is an example of a useful
constraint that is convex.  It states that the sum of the $K$ largest
absolute weights cannot exceed some limit.  As a specific example,
we can require that no collection of five assets can have a total
absolute weight of more than 30\%~\cite{Perrin2020,aps2023mosek}.
A \emph{minimum nonzero holding} constraint is an example of a commonly
imposed nonconvex constraint.
It states that any nonzero weight must have an absolute
value exceeding some given minimum, such as $0.5\%$.
(This one is easily handled using a heuristic based on convex optimization;
see \S\ref{s-nonconvex}.)

\subsection{Trades}

\paragraph{Trade vector.}
We let $w^\text{pre}$ and $c^\text{pre}$
denote the pre-trade portfolio weights, \ie, the
portfolio weights before we carry out the trades to construct the portfolio
given by $w$ and $c$.
We need the pre-trade weights to account for transaction costs.
We refer to
\BEQ\label{e-z-def}
z= w-w^\text{pre},
\EEQ
the current weights minus the previous ones,
as the (vector of) \emph{trades} or the \emph{trade list}.
These trades have a simple interpretation: $z_i = 0.01$ means
we buy an amount of asset $i$ equal in value to 1\% of our
total portfolio value, and $z_i=-0.03$ means we sell an amount of asset
$i$ equal to $3\%$ of the portfolio value.

Since $\ones^T w^\text{pre}+ c^\text{pre}=1$, we have
\BEQ\label{e-c-post-trade}
c = c^\text{pre}- \ones^T z,
\EEQ
\ie, the post-trade cash weight is the pre-trade cash weight minus the net
weight of the trades.
This does not include holding and transaction costs, discussed below.

\paragraph{Turnover.}
The quantity
\[
T = \frac{1}{2} \sum_{i=1}^n |z_i| = \frac{1}{2} \| z\|_1
\]
is the \emph{turnover}.  Here too, several other different but closely related
definitions are also used, for example the minimum
of the total weight bought and the total weight sold \cite[Chap.~16]{grinold2000active}.
A turnover $T=0.01$ means that the average of total amount bought and total amount sold
is 1\% of the total portfolio value.
The turnover is often annualized, by multiplying by the number of
trading periods per year.

\subsection{Trading constraints and costs}

We typically have constraints on the trade vector $z$, as well
as a trading cost that depends on $z$.

\paragraph{Trade limits.}
Trade limits impose lower and upper bounds on trades, as
\[
z^\text{min} \leq z \leq z^\text{max},
\]
where $z^\text{min}$ and $z^\text{max}$ are given limits.
These trade limits can be used to limit market participation,
defined as the ratio of the magnitude of each trade to the
trading volume, using, \eg,
\BEQ\label{e-z-limit}
|z| \leq 0.05 v,
\EEQ
where $v\in \reals^n$ is the trading volumes of
the assets, expressed as multiples of the portfolio value.
This constraint limits our participation for each asset to be
less than 5\%.
(It corresponds to trade limits $z^\text{max} = - z^\text{min} = 0.05v$.)
Since the trading volumes are not known when $z$ is chosen, we
use a forecast instead of the realized trading volumes.

\paragraph{Turnover limit.}
In addition to trade limits, we can limit the turnover as
\BEQ\label{e-turnover-limit}
T \leq T^\text{tar},
\EEQ
where $T^\text{tar}$ is a specified turnover limit.

\paragraph{Trading cost.}
Trading cost refers to the cost of carrying out a trade.
For example, if we buy a small quantity of an asset,
we pay the ask price, while if we sell an asset,
we receive the bid price. Since the nominal price of an asset
is the midpoint between the ask and bid prices, we can think
of buying or selling the asset as doing so at the nominal price,
plus an additional positive cost that is the trade amount times
one-half the bid-ask spread.
This \emph{bid-ask spread transaction cost} has the form
\[
\sum_{i=1}^n \kappa^\text{spread}_i |z_i| = (\kappa^\text{spread})^T |z|,
\]
where $\kappa^\text{spread}\in \reals^n$ is the vector of one-half the
asset bid-ask spreads (which are all positive).
This is the transaction cost expressed as a fraction of the portfolio value.
For small trades this is a reasonable approximation of transaction cost.

For larger trades we `eat through' the order book.
To buy a quantity of an asset,
we buy each ask lot, in order from lowest price,
until we fill our order. An analogous situation occurs when selling.
This means that we end up paying more per share than the
ask price when buying, or receiving less than the bid price
when selling. This phenomenon is called \emph{market impact}.

A useful approximation of transaction cost that takes
market impact into account is
\BEQ\label{e-t-cost}
\phi^\text{trade} (z) = (\kappa^\text{spread})^T|z| +
(\kappa^\text{impact})^T |z|^{3/2},
\EEQ
where the first term is the bid-ask spread component
of transaction cost, and the second models the market impact, \ie,
the additional cost incurred as the trade eats through the order book.
The vector $\kappa^\text{impact}$ has positive entries and typically takes
the form
\[
\kappa^\text{impact}_i = a s_i v_i^{-1/2},
\]
where $s_i$ is the volatility of asset $i$ over the trading period, $v_i$ is the
volume of market trading, expressed as a multiple of the portfolio value, and
$a$ is a constant on the order of one; see \cite{grinold2000active,
loeb1983trading, toth2011anomalous, mastromatteo2014impact}
and \cite[\S2.3]{BoydKahnMultiPeriod}. Evidently the transaction cost increases
with volatility, and decreases with market volume. Several other approximations
of transaction cost are used~\cite{almgren2000optimal, robert2012measuring}.

\paragraph{Liquidation cost.}
Suppose we liquidate the portfolio, \ie, close out all asset positions,
which corresponds to the trade vector $z=-w$. The \emph{liquidation cost} is
\[
\phi^\text{trade}(-w) = (\kappa^\text{spread})^T |w| +
(\kappa^\text{impact})^T |w|^{3/2}.
\]
If the liquidation is carried out over multiple periods, the
bid-ask term stays the same, but the market impact term decreases.
For this reason a common approximation of the liquidation cost
ignores the market impact term.
A liquidation cost constraint has the form
\BEQ\label{e-l-cost-constr}
(\kappa^\text{spread})^T |w| \leq \ell ^\text{max},
\EEQ
where $\ell^\text{max}$ is a maximum allowable liquidation cost,
such as $1\%$.
This is a weight constraint; it limits our holdings in
less liquid assets, which have higher bid-ask spreads.
It can be interpreted as a liquidity-weighted leverage
(taking the bid-ask spread as a proxy for liquidity).
When all assets have the same bid-ask spread, the liquidation constraint
reduces to a leverage constraint. For example with all bid-ask spreads
equal to $0.001$ (\ie, 10 basis points or bps)
and a maximum liquidation cost $\ell^\text{max} = 0.01$ (\ie, 1\% of the total
portfolio value),
the liquidation cost limit \eqref{e-l-cost-constr}
reduces to a leverage limit \eqref{e-leverage-limit} with $L^\text{tar}=10$.

\paragraph{Transaction cost forecasts.}
When the trades $z$ are chosen, we do not know the bid-ask spreads,
the volatilities, or the volumes.  Instead we use forecasts of these
quantities in \eqref{e-z-limit}, \eqref{e-t-cost}, and
\eqref{e-l-cost-constr}.
Simple forecasts, such as a trailing average or median of realized
values, are typically used.
More sophisticated forecasts take can into account calendar effects
such as seasonality, or the typically low trading volume the day after
Thanksgiving.

\section{Return and risk forecasts} \label{s-return-and-risk}

\subsection{Return}
\paragraph{Gross portfolio return.}
We let $r_i$ denote the return, adjusted for dividends, splits, and other corporate
actions, of asset $i$ over the investment period.
We collect these asset returns into a return vector
$r=(r_1, \ldots, r_n)\in \reals^n$. The portfolio return from asset $i$ is $r_i w_i$.
We let $r^\mathrm{rf}$ denote the risk-free interest rate, so the return in
the cash account is $r^\mathrm{rf}c$.
The (gross) total portfolio return is then
\[
R = r^T w + r^\mathrm{rf} c.
\]
This gross return does not include holding or trading costs.
A closely related quantity is the \emph{excess return}, the portfolio return minus
the risk-free return,
$R-r^\mathrm{rf} = r^Tw + r^\text{rf}(c-1)$.

\paragraph{Net portfolio return.}
The net portfolio return is the gross return minus the holding costs and
transaction costs,
\BEQ\label{e-net-return}
R^{\textrm{net}} = R - \phi^{\text{hold}}(w) - \phi^{\text{trade}}(z).
\EEQ

\paragraph{Active return.}
The \emph{active portfolio return}
is the return relative to a benchmark portfolio,
\[
r^T w + r^\mathrm{rf} c - r^T w^\mathrm{b} = r^T (w-w^\mathrm{b}) + r^\mathrm{rf} c.
\]
If we subtract holding and trading costs we obtain the
\emph{net active portfolio return}.

\paragraph{Cash as slack.}
Since we do not know but only forecast the bid-ask spread, volatility,
and volume, which appear in the transaction cost \eqref{e-t-cost}
(which is itself only an approximation) we should consider the post-trade cash
$c$ in \eqref{e-c-post-trade} as an approximation that uses a
forecast of holding and transaction costs, not
the realized holding and transaction costs.
We do not expect the realized post-trade cash weight to be exactly $c$.

\subsection{Probabilistic asset return model}
When we choose the trades $z$ we do not know the asset returns $r$.
Instead, we model $r$ as a multivariate random variable with mean
$\mu\in \reals^n$ and covariance matrix $\Sigma \in \symm_{++}^n$ (the
set of symmetric positive definite $n \times n$ matrices),
\[
\Expect r = \mu, \qquad \Expect (r-\mu)(r-\mu)^T = \Sigma.
\]
The entries of the mean $\mu$ are often referred to as
\emph{trading  signals}~\cite{isichenko2021quantitative}.
The asset return mean and covariance are forecasts, as described below.
The asset return volatilities $s\in \reals^n$ appearing in the
transaction cost model \eqref{e-t-cost} can be expressed
as $s=\diag(\Sigma)^{1/2}$, where the squareroot is elementwise.

\paragraph{Expected return and risk.}
With this statistical model of $r$, the portfolio return $R$ is a
random variable with mean $\bar R = \Expect R$
and variance $\sigma^2 = \var R$ given by
\[
\bar R = \mu^T w + r^\mathrm{rf} c, \qquad
\sigma^2 = w^T\Sigma w.
\]
The \emph{risk} of the portfolio is defined as the standard
deviation of the portfolio return, \ie, $\sigma$.

Similarly, the active return $R^\text{a}$ is a random variable with
mean and variance
\[
\bar R^\text{a} = \mu^T (w-w^\text{b}) + r^\mathrm{rf} c = \bar R -
\mu^T w^\text{b}, \qquad
(\sigma^\text{a})^2 = (w-w^\text{b})^T \Sigma (w-w^\text{b}),
\]
and the \emph{active risk} is $\sigma^\text{a}$.
The risk and active risk are often given in annualized form, obtained by
multiplying them by the squareroot of the number of periods per year.

The parameters $\mu$ and $\Sigma$ are estimates or forecasts of
the statistical model of asset returns, which is itself an approximation.
For this reason the risk $\sigma$ is
called the \emph{ex-ante} risk, to distinguish it from the
standard deviation of the realized portfolio returns when trading,
the \emph{ex-post} risk.
Similarly we refer to $\sigma^\text{a}$ as the ex-ante active risk.

\paragraph{Optimizing expected return and risk.}
We have two objectives, high expected return and low risk.
Perhaps the most common method for combining these objectives is to form a \emph{risk-adjusted return},
\[
\bar R - \gamma \sigma^2,
\]
where $\gamma>0$ is the \emph{risk aversion parameter}.
Maximizing risk-adjusted return (possibly with other objective terms,
and subject to constraints) gives the desired portfolio.
Increasing $\gamma$ gives us a portfolio with lower risk and also lower
expected return.  The risk aversion parameter allows us to explore the
risk-return trade-off.
This risk-adjusted return approach became popular in part because
the resulting optimization problem is typically a quadratic program (QP),
for which reliable solvers were developed even in the 1960s.

Another approach is to maximize expected return
(possibly with other objective terms), subject to a
\emph{risk budget} or \emph{risk target} constraint
\BEQ\label{e-risk-limit}
\sigma \leq \sigma^\text{tar}.
\EEQ
(The corresponding optimization problem is not a QP, but is
readily handled by convex optimization solvers developed
in the 1990s \cite{lobo1998applications,nesterov1994interior,sturm1999using,toh1999sdpt3}.)
This formulation seems more natural, since a portfolio manager
will often have a target risk in her mind, \eg, 8\% annualized.
This is the basic formulation that we recommend.

There are many other ways to combine expected return and risk.
For example, we can maximize the return/risk ratio, called the \emph{Sharpe
ratio} (with no benchmark) or
\emph{information ratio} (with a benchmark).
This problem too can be solved via convex optimization, at least
when the constraints are simple~\cite{cvx_book_additional}.

\subsection{Factor model}\label{s-factor-model}
In practice, and especially for large universes,
it is common to use a \emph{factor model} for the returns.
The factor return model, with $k$ factors (typically with $k\ll n$), has the form
\BEQ\label{e-factor-model}
r = F f + \epsilon,
\EEQ
where $F \in \reals^{n \times k}$ is the \emph{factor loading matrix},
$f \in \reals^k$ is the vector of \emph{factor returns},
and $\epsilon \in \reals^n$ is the \emph{idiosyncratic return}.
The term $Ff$ is interpreted as the component of asset returns explainable
or predicted by the factor returns.

At portfolio construction time the factor loading matrix $F$ is known,
and the factor return $f$ and idiosyncratic return $\epsilon$
are modeled as uncorrelated random variables with means and covariance
matrices
\[
\Expect f = \bar f, \qquad \cov f = \Sigma^\text{f}, \qquad
\Expect \epsilon = \bar \epsilon, \qquad \cov \epsilon = D,
\]
where $D$ is diagonal (with positive entries).  The entries $\bar \epsilon$,
the means of the idiosyncratic returns, are also referred to as the \emph{alphas},
especially when there is only one factor which is the overall market return.
They are the part of the asset returns not explained by the factor returns.

With the factor model \eqref{e-factor-model} the asset
return mean and covariance are
\[
\mu = F \bar f + \bar \epsilon, \qquad
\Sigma = F \Sigma^f F^T + D.
\]
The return covariance matrix in a factor model has a special form,
low rank plus diagonal.
The portfolio return mean and variance are
\[
\bar R = (F \bar f)^Tw + \bar \epsilon^Tw + r^{\text{rf}}c, \qquad
\sigma^2 = (F^Tw)^T \Sigma^\text{f}(F^Tw) + w^T D w.
\]

The factor returns are constructed to have explanatory power for
the returns of assets in our universe.
For equities, they are typically the returns of other portfolios, such as the 
overall market (with weights proportional to capitalization), industries, 
and style portfolios like the celebrated
Fama-French factors~\cite{fama1992cross, fama1993common}.
For bonds, the factors are typically constructed from
yield curves, interest rates, and spreads.
These traditional factors are interpretable.

Factors can also can be constructed directly from previous realized
asset returns using methods such as principal component analysis 
(PCA)~\cite{bai2008large, bai2003inferential,
lettau2020estimating, lettau2020factors, pelger2022factor,
pelger2022interpretable}.  Aside from the first principal component, which typically
is close to the market return, these factors are less interpretable.

\paragraph{Factor and idiosyncratic returns.}
A factor model gives an alternative method to specify the 
expected return as $\mu = F\bar f + \bar \epsilon$,
where $\bar f$ is a forecast of the 
factor returns and $\bar \epsilon$ is a forecast of the 
idiosyncratic returns, \ie, the asset alphas.
One common method uses only a forecast of the factor returns, 
with $\bar \epsilon=0$, so $\mu=F\bar f$.
A complementary method assumes zero factor returns, 
so we have $\mu=\bar \epsilon$, \ie, the mean asset 
returns are the same as the idiosyncratic asset mean returns.

\paragraph{Factor betas and neutrality.}
Under the factor model \eqref{e-factor-model}, the covariance of
the portfolio return $R$ with the factor returns $f$ is the $k$-vector
\[
\cov (R,f) = \Sigma^\text{f} F^Tw.
\]
The \emph{betas} of the portfolio with respect to the factors
divide these covariances by the variance of the factors,
\[
\beta = \diag(s^\text{f})^{-2}\Sigma^\text{f} F^Tw,
\]
where $s^\text{f} = \diag(\Sigma^\text{f})^{-1/2}$ is the
vector of factor return volatilities.

The constraint that our portfolio return is
uncorrelated (or has zero beta) with the $i$th factor return $f_i$, 
under the factor model \eqref{e-factor-model}, is
\BEQ\label{e-factor-neutrality}
\cov (R,f)_i= (\Sigma^f F^T w)_i =0.
\EEQ
This is referred to as \emph{factor neutrality}
(with respect to the $i$th factor).
It is a simple linear equality constraint, which can be
expressed as $a^T w = 0$, where $a$ is the $i$th column of 
$F\Sigma^\text{f}$.
Factor neutrality constraints are typically used with active weights.  
In this case, factor neutrality means that the portfolio beta matches the 
benchmark beta for that factor. This also is a linear equality constraint 
that can be expressed as $\beta_i=\beta^\text{b}_i$, with $\beta^\text{b}$
the benchmark betas.

\paragraph{Advantages of a factor model.}
Especially with large universes, the factor model
(specified by $F$, $\Sigma^\text{f}$, and $D$) can give a
better estimate of the return covariance, compared to methods that directly
estimate the $n \times n$ matrix $\Sigma$~\cite{johansson2023covariance}.
Another substantial advantage is computational.  By exploiting the
low-rank plus diagonal structure of the return covariance with a factor model,
we can reduce the
computational complexity of solving the Markowitz optimization problem
from $O(n^3)$ flops (without exploiting the factor model) to $O(nk^2)$
flops (exploiting the factor form).
These computational savings can be dramatic, \eg, for a whole world portfolio
with $n=10000$ and $k=100$, where we obtain a $10000$ fold decrease in
solve time; see \S\ref{s-scaling}.

\subsection{Return and risk forecasts}
Here we briefly discuss the forecasting of $\mu$ and $\Sigma$
(or $F$, $\Sigma^\text{f}$, and $D$ in a factor model).
Markowitz himself did not address the question of 
estimating $\mu$ and $\Sigma$;
when asked by practitioners how one should choose these forecasts,
his reply was~\cite{siegel2023}
\begin{quote}
\centering
\enquote{\emph{That's your job, not mine.}}
\end{quote}
It is well documented that poor or na\"{i}ve estimates of these, \eg, 
the sample mean and covariance, can yield poor
portfolio performance~\cite{michaud1989markowitz_enigma}. 
But even reasonable forecasts will have errors, which can degrade 
performance.  We show some methods to mitigate this 
forecast uncertainty in \S\ref{s-robust}.

\paragraph{Asset returns estimate.}
The expected returns vector $\mu$ is by far the
most important parameter in the portfolio construction process,
and methods for estimating it, or the factor and idiosyncratic 
return means are for obvious reasons in general
proprietary. It is also the most challenging data to estimate. There is no
consensus on how to estimate the mean returns, and the literature is vast. 

Regularization methods can improve mean estimates.
As an example, the Black-Litterman model~\cite{black1990asset}
allows a portfolio manager to incorporate her views on how the expected
returns differ from the market consensus, and in essence acts as a form of
regularization of the portfolio toward the market portfolio.
Another method that serves implicitly as regularization is winsorization,
where the mean estimates are clipped when they go outside a specified
range~\cite{WelschWinsorization2007}, \cite[Chap. 14]{grinold2000active}.
Yet another method is cross-sectionalization,
where the preliminary estimate of returns $\mu$ is replaced with $\tilde \mu$,
the same values monotonically mapped to (approximately) a Gaussian distribution
\cite[Chap. 14]{grinold2000active}.

\paragraph{Return covariance estimate.}
There are many ways to estimate the covariance
matrix, with or without a factor model.
Approaches that work well in practice include the exponentially
weighted moving average (EWMA)~\cite{ongerstaey1996riskmetrics}, dynamic
conditional correlation (DCC)~\cite{DCC}, and iterated 
EWMA~\cite{cov_barrat_2022}. For a detailed discussion on how to 
estimate a covariance matrix for
financial return data, see~\cite{johansson2023covariance} and the 
references therein.

\subsection{Making return and risk forecasts robust}\label{s-robust}
In this section we address methods to mitigate
the impact of forecast errors in return and covariance
estimation, which can lead to poor performance.
This directly addresses one of the main criticisms of the Markowitz method,
that it is too sensitive to estimation errors.
Here, we briefly review how to address robust return mean and covariance
estimation, and refer the reader to~\cite[\S4.3]{BoydKahnMultiPeriod}
and~\cite{fabozzi2007robust, tutuncu2004robust} for more
detailed discussions.

\paragraph{Robust return forecast.}
We model our uncertainty in the mean return vector by giving an interval
of possible values for each return mean.  
We let $\mu\in \reals^n$ denote our nominal
estimate of the return means, and we take the nonnegative
vector $\rho \in \reals^n_+$ to describe the half-width or radius of the
uncertainty intervals.
Thus we imagine that the return can be any vector of the form
$\mu + \delta$, where $|\delta_i| \leq \rho_i$.
For example $\mu_i = -0.0010$ and $\rho_i =0.0005$ means that
the mean return for asset $i$ lies in the range $[-15,-5]$ bps.

We define the \emph{worst-case mean portfolio return} as the
minimum possible mean portfolio return consistent with the given ranges of
asset return means:
\[
R^{\text{wc}} = \min \{ (\mu + \delta)^T w \mid |\delta| \leq \rho \}.
\]
We can think of this as an adversarial game. The portfolio manager (PM) chooses
the portfolio $w$, and an adversary then chooses the worst mean return 
consistent with the given uncertainty intervals.
This second step has an obvious solution: We choose $\mu_i-\rho_i$
when $w_i\geq 0$, and we choose $\mu_i+\rho_i$ when $w_i<0$.
In words: For long positions the worst return is the minimum possible;
for short positions the worst return is the maximum possible.
With this observation, we obtain a simple formula for the worst-case
portfolio mean return,
\BEQ\label{e-wc-portfolio-return}
R^{\text{wc}} = \bar R - \rho^T | w |.
\EEQ
The first term is the nominal mean return; the second term, which is
nonpositive, gives the degradation of return induced by the uncertainty.
We refer to $\rho^T|w|$ as the portfolio \emph{return forecast error
penalty}
in our return forecast.
The return forecast error penalty has a nice interpretation
as an uncertainty-weighted leverage.

When the portfolio is long-only, so $w\geq 0$, the worst-case
asset returns are obvious: they simply take their minimum values, $\mu-\rho$.
In this case the worst-case portfolio mean return
\eqref{e-wc-portfolio-return} is the usual
mean portfolio return, with each nominal asset return reduced by its
uncertainty.

The return forecast uncertainties $\rho$ can be chosen by several methods.
One simple method is to set all entries the same, 
and equal to some quantile of the entries of $|\mu|$, such as the 20th
percentile.  A more sophisticated method relies on multiple forecasts of the returns,
and sets $\mu$ as the mean or median forecast, and $\rho$ as some measure 
of spread, such as standard deviation, of the forecasts.

\paragraph{Robust covariance forecast.}
We can also consider uncertainty in the covariance matrix.  We let $\Sigma$
denote our nominal estimate of the covariance matrix.
We imagine that the covariance matrix has the form
$\Sigma + \Delta$ where $\Delta\in\symm^n$ (the set of symmetric $n \times n$
matrices) where the perturbation $\Delta$ satisfies
\[
|\Delta_{ij}| \leq \varrho (\Sigma_{ii}\Sigma_{jj})^{1/2},
\]
where $\varrho \in [0,1)$ defines the level of uncertainty.
For example, $\varrho = 0.04$ means that the diagonal elements of
the covariance matrix can change by up to 4\%
(so the volatilites can change by around 2\%),
and the asset return correlations can change by up to around $4\%$.
(You should not trust anyone who claims that his
asset return covariance matrix estimate is more accurate than this.)

We define the \emph{worst-case portfolio risk} as
the maximum possible risk over covariance matrices consistent with our
uncertainty set,
\[
\left( \sigma^{\text{wc}} \right)^2 =
\max \{ w^T(\Sigma + \Delta)w \mid
|\Delta_{ij}| \leq \varrho (\Sigma_{ii}\Sigma_{jj})^{1/2} \}.
\]
This can be expressed analytically as \cite[\S4.3]{BoydKahnMultiPeriod}
\BEQ\label{e-wc-portfolio-risk}
\left( \sigma^{\text{wc}} \right)^2 =
\sigma^2 + \varrho \left(\sum_{i=1}^n \Sigma_{ii}^{1/2}|w_i|\right)^2.
\EEQ
The second term is the \emph{covariance forecast error penalty}.
It has a nice interpretation as an additive regularization term,
the square of a volatility-weighted leverage.
The worst-case risk can be expressed using Euclidean norms as
\BEQ\label{e-wc-risk-norm}
\sigma^{\text{wc}} = \left\| \left(
\sigma, \sqrt{\varrho} (\diag(\Sigma)^{1/2})^T |w_i|
\right) \right\|_2.
\EEQ

When the portfolio is long-only, the worst-case risk
\eqref{e-wc-portfolio-risk} can be simplified.
In this case, the worst-case risk is the risk using the covariance
matrix $\Sigma + \varrho ss^T$, where $s=\diag(\Sigma)^{1/2}$ is
vector of asset volatilities, under the nominal covariance.

\section{Convex optimization formulation}\label{s-cvx-formulation}

\subsection{Markowitz problem}\label{s-markowitz-problem}
In this section we assemble the objective terms and constraints
described in \S\ref{s-portfolio-and-holdings} and
\S\ref{s-return-and-risk}
into one convex optimization problem.
We obtain the Markowitz problem
\BEQ\label{e-markowitz-hard}
\begin{array}{ll}
\mbox{maximize} & R^\text{wc} -
\gamma^\text{hold}\phi^\text{hold}(w,c) -
\gamma^\text{trade}\phi^\text{trade}(z)\\
\mbox{subject to} &  \ones^T w + c = 1, \quad z=w-w^\text{pre},\\
                   &  w^\text{min} \leq w \leq w^\text{max},
		 \quad L \leq L^\text{tar}, \quad
                   c^\text{min} \leq c \leq c^\text{max},\\
                   &  z^\text{min} \leq z \leq z^\text{max},
		 \quad T \leq T^\text{tar},\\
                   &  \sigma^\text{wc} \leq \sigma^\text{tar},
\end{array}
\EEQ
with variables $w\in \reals^n$  and $c\in \reals$,
and positive parameters $\gamma^\text{hold}$ and
$\gamma^\text{trade}$ that allow us to scale the holding and transaction costs,
respectively.
Despite the nonlinear and nondifferentiable functions appearing
in the objective and constraints, this is a convex optimization problem,
which can be very reliably and efficiently solved.
We can add other convex objective terms to this problem, such as
factor neutrality or liquidation cost limit, or work with
active risk and return with a benchmark.

The objective is our forecast of the (robustified, worst-case)
net portfolio return, with
the holding and transaction costs scaled by the
parameters $\gamma^\text{hold}$ and $\gamma^\text{trade}$,
respectively.
The first line of constraints relate the pre-trade portfolio,
which is given, and the post-trade portfolio, which is to be chosen.
The second line of constraints are weight limits, and the
third line contains the trading constraints.
The last line of constraints is the (robustified, worst-case) risk limit.

\paragraph{Data.}
We divide the constants that need to be specified in the problem
\eqref{e-markowitz-hard} into two groups,
\emph{data} and \emph{parameters},
although the distinction is not sharp.
Data are quantities we \emph{observe} (such as the previous portfolio weights)
or forecast (such as return means, market volumes, or bid-ask spreads):
\BIT
\item \emph{Pre-trade portfolio weights} $w^\text{pre}$ and $c^\text{pre}$.
\item \emph{Asset return forecast} $\mu$.
\item \emph{Risk model} $\Sigma$, or
for a factor model, $F$, $\Sigma^\text{f}$, and $D$.
\item \emph{Holding cost parameters} $\kappa^\text{short}$ and $\kappa^\text{borrow}$.
\item \emph{Trading cost parameters}
$\kappa^\text{spread}$ and
$\kappa^\text{impact}$ (which in turn
depend on the forecast bid-ask spreads, asset volatilities, and market volume).
\EIT

\paragraph{Parameters.}
Parameters are quantities that we \emph{choose} in order to obtain good
investment performance, or to reflect portfolio manager preferences,
or to comply with legal requirements or regulations.  These are
\BIT
\item \emph{Target risk} $\sigma^\text{tar}$.
\item \emph{Holding and trading scale factors}
$\gamma^\text{hold}$ and $\gamma^\text{trade}$.
\item \emph{Weight and leverage limits}
$w^\text{min}$,
$w^\text{max}$,
$c^\text{min}$,
$c^\text{max}$,
and $L^\text{tar}$.
\item \emph{Trade and turnover limits}
$z^\text{min}$,
$z^\text{max}$,
and $T^\text{tar}$.
\item \emph{Mean and covariance forecast uncertainties}
$\rho$ and $\varrho$.
\EIT
We list the mean and covariance forecast uncertainties as parameters
since they are closer to being chosen than measured or estimated.
When the mean return uncertainties are chosen as described above from
a collection of return forecasts, they would be closer to data.

\paragraph{Initial default choices for parameters.}
The target risk, and the weight, leverage, trade, and turnover limits 
are interpretable and can be assigned reasonable values by the PM.  
The return and risk uncertainty parameters $\rho$ and $\varrho$ can be 
chosen as described above.
The hold and trade scale factors can be chosen to be around one.

To improve performance the PM will want to adjust or tune
these parameter values around their natural or default values,
as discussed in \S\ref{s-tuning}.

\subsection{Softening constraints} \label{s-soft-constr}
The Markowitz problem \eqref{e-markowitz-hard}
includes a number of constraints.
This can present two challenges in practice.  First, it can lead to
substantial trading, for example to satisfy our leverage or 
ex-ante worst-case risk limits, even when they would have been 
violated only slightly, which can lead
to poor performance due to excessive trading.
Second, the problem can be infeasible, meaning there is
no choice of the variables that satisfy all the constraints.
This can complicate back-tests or simulations, as well as
running the trading policy in production, where such infeasiblilities 
naturally occur most frequently during periods of market 
stress, putting the PM under additional pressure.

\paragraph{Soft constraints.}
Here we explain a standard method in optimization, in which
some of the constraints can be softened,
which means we allow them to be somewhat violated, if needed.
In optimization, softness refers to how much we care about
different values of an objective.
We can think of the objective as infinitely soft: We will accept any
objective value, but we prefer larger values (if we are maximizing).
We can think of constraints as infinitely hard: We will not accept
any violation of them, even if it is only by a small amount.
\emph{Soft constraints}, described below, are in between.  
They should normally act as constraints, but when needed, 
they can be violated.
When a soft constraint is violated, and by how much, depends on our priorities,
with high priority meaning that the constraint should be violated only
when absolutely necessary.

Consider a (hard) constraint such as $f \leq f^\text{max}$.
This means that we will not accept any choice of the variables for
which $f>f^\text{max}$.
To make it a \emph{soft constraint}, we remove the constraint from the problem
and form a penalty term
\[
\gamma (f-f^\text{max})_+
\]
which we subtract from the objective, when we are maximizing.
The number $(f-f^\text{max})_+$ is the \emph{violation} of the original
constraint $f \leq f^\text{max}$.
The positive parameter $\gamma$ is called the \emph{priority
parameter} associated with the softened constraint.
In this context, we refer to the parameter $f^\text{max}$ as a
\emph{target} for the value of $f$, not a limit.
With the softened constraint, we can accept variable choices
for which $f> f^\text{max}$, but the optimizer tries to avoid this given
the penalty paid (in the objective) when this occurs.
Softening constraints preserves convexity of a problem.

\paragraph{Markowitz problem with soft constraints.}
A number of constraints in \eqref{e-markowitz-hard}
should be left as (hard) constraints.
These include the constraints relating the proposed and previous
weights, \ie, the first line of constraints in
\eqref{e-markowitz-hard}.
When the portfolio is long-only, the constraint $w\geq 0$
should be left as a hard constraint, and similarly for a
constraint such as $c \geq 0$, \ie, that we do not borrow cash.
When a leverage limit is strict or imposed by a mandate,
it should be left as a hard constraint; when it is imposed by the portfolio
manager to improve performance, or more likely, to help her avoid
poor outcomes, it can be softened.

The other constraints in \eqref{e-markowitz-hard} are candidates
for softening.  Weight and trade limits, including leverage
and turnover limits, should be softened (except in the cases
described above).
The worst-case risk limit $\sigma \leq \sigma^\text{tar}$ should be softened,
with a risk penalty term 
\[
\gamma^\text{risk}(\sigma-\sigma^\text{tar})_+
\]
subtracted from the objective.
When the associated priority parameter $\gamma^\text{risk}$ is chosen
appropriately, this allows us to occasionally violate our risk limit a bit
when the violation is small.
We refer to the softened Markowitz problem as the Markowitz++ problem.

One nice attribute of the Markowitz++ problem is that it is always
feasible; the choice $z=0$, \ie, no trading, is always feasible, 
even when it is a poor choice.  
This means that the softened Markowitz problem can be
used to define a trading policy
that runs with little or no human intervention (with, however,
any soft constraints that exceed their targets reported to the portfolio
manager).

\paragraph{Priority parameters.}
When we soften the worst-case risk, leverage, and turnover constraints, we
gain several more parameters,
\[
\gamma^\text{risk}, \quad
\gamma^\text{lev}, \quad
\gamma^\text{turn}
.
\]
Evidently the larger each of these priority parameters is, the more reluctant
the optimizer is to violate it.  (Here we anthropomorphize the optimization 
problem solver.)
When the priority parameters are large, the associated
soft constraints are effectively hard.
Beyond these observations, however, it is hard to know what values should be used.

\paragraph{Choosing priority parameters.}\label{p-choosing-priority-paramns}
Here we describe a simple method to obtain reasonable useful initial values
for the priority parameters associated with softened constraints.
Our method is based on Lagrange multipliers or dual variables.
Suppose we solve a problem with hard constraints, and obtain optimal
Lagrange multipliers for each of the constraints. If we use these Lagrange
multipliers as priorities in a softened version of the problem, 
all the original constraints will be satisfied.  Roughly speaking, the Lagrange
multipliers give us values of priorities for which the soft constraints
are effectively hard.  We would want to use priority values a bit smaller, so
that the original constraints can occasionally be violated.

Now we describe the method in detail.
We start by solving multiple instances of the problem with hard constraints,
for example in a back-test, recording the values of the Lagrange multipliers
for each problem instance (when the problem is feasible).
We then set the priority parameters to some quantile, such as
the 80th percentile, of the Lagrange multipliers.
With this choice of priority parameters, we expect (very roughly) 
the original constraints to hold around 80\% of the time.
For hard constraints that are only occasionally tight, another method for 
choosing the priority parameters is as a fraction of the maximum Lagrange
multiplier observed.

Using this method we can obtain reasonable starting values of the priority parameters.
The final choice of priority parameters is done by back-testing and parameter
tuning, starting from these reasonable values, as discussed in \S\ref{s-backtest}.

\subsection{Nonconvex constraints and objectives}\label{s-nonconvex}
All objective terms and constraints discussed so far are convex,
and the Markowitz problem \eqref{e-markowitz-hard}, and its 
softened version, are convex optimization problems.
They can be reliably and efficiently solved.

Some other constraints and objective terms are not convex. 
The most obvious one is that the trades must ultimately involve
an integer number of shares.
As a few other
practical examples, we might limit the number of nonzero weights,
or insist on a minimum nonzero weight absolute value. 
When these constraints are added to \eqref{e-markowitz-hard}, the problem
becomes nonconvex.
Great advances have been made in solvers that handle so-called
mixed-integer convex problems \cite{Kronqvist2018}, and these can be used to 
solve these portfolio construction problems.
The disadvantage is longer solve time,
compared to a similar convex problem, and sometimes, dramatically 
longer solve time if we insist on solving the problem to global optimality.
A convex portfolio construction problem that can be solved
in a small fraction of a second can take many seconds, or even minutes or more,
to solve when nonconvex constraints are added.

For production, where the problem is solved daily, or even hourly, this is fine.
The slowdown incurred with nonconvex optimization is however very bad for 
back-testing and validation, where many thousands,
or hundreds of thousands, of portfolio construction problems are to be solved.
One sensible approach is to carry out back-testing using a convex formulation,
so as to retain the speed and reliability of a convex optimization, and 
run a nonconvex version in production.
As a variant on this, back-tests using convex optimization can be used
for parameter search, and one final back-test with a nonconvex formulation
can be used to be sure the results are close.
Running backtests using only convex constraints works
because the nonconvex constraints typically only have a small impact on the portfolio
and its performance.

\paragraph{Heuristics based on convex optimization.}
Essentially all solvers for nonconvex problems that attempt to find a global
solution rely on convex optimization under the hood \cite{horst2013global}.
The issue is that a very large number of convex optimization problems
might need to be solved to find a global solution.

But many nonconvex constraints can be handled heuristically by solving just a 
few convex optimization problem.
As a simple example we might simply round the numbers of 
shares in a trade list to an integer.  This rounding should have little effect
unless the portfolio value is very small.

Other nonconvex constraints are readily handled by heuristics that 
involve solving just a handful of convex problems.
One general method is called relax-round-solve \cite{NCVX}.
We illustrate this method to handle the constraint that the 
minimum nonzero weight absolute value is $0.001$ (10 bps).
First we solve the problem ignoring this constraint.  If the weights satisfy
the constraint, we are done (and the choice is optimal).  If not,
we set a threshold and divide the assets into
those with absolute weight smaller than the threshold,
those with weights larger than the threshold,
and those which are less than minus the threshold.
We then add constraints to the original problem, setting the weights
to zero, more than $0.001$, and less than $-0.001$, depending on
the weights found in the first problem.  These are convex constraints, and when we solve the
second time we are guaranteed to satisfy the nonconvex constraint.
We thus solve two convex problems.  In the first one, we essentially decide 
which weights will be zero, which will be more than the minimum nonzero long
weight, and which will be short more than the minimum.
In the second one we adjust all the weights, ensuring that the minimum
absolute nonzero weight constraint holds.

\subsection{Back-testing and parameter tuning}\label{s-backtest}
\paragraph{Back-testing.}
Back-testing refers to simulating a trading strategy using historical data.
To do this we provide the forecasts for all quantities needed, 
including the mean return and covariance, for Markowitz portfolio
construction in each period.
In each period these forecasts, together with the parameters, 
are sent in to the Markowitz portfolio
construction method, which determines a set of trades.
We then use the \emph{realized} values of return, volatility, 
bid-ask spread, and market volume to compute the 
(simulated) realized net return
$R^\text{net}_t$, where the subscript gives the time period.
Note that while the Markowitz trading engine uses forecasts of 
various quantities, the simulation uses the realized historical values.
This gives a reasonably realistic approximation of what the
result would have been, had we actually carried out this trading.
(It is still only an approximation, since it uses our particular
trading cost model.  Of course a more complex or realistic trading 
cost model could be used for simulation.)
The back-test simulation can also include practical aspects like 
trading only an integer number of shares or blocks of shares.
The simulation can also include external cash entering or leaving the 
portfolio, such as liabilities that must be paid each period.

In the simulation we log the trajectory of the portfolio.  We can compute
various quantities of interest such as the realized return, volatility, Sharpe 
or information ratio, turnover, and leverage, all potentially over
multiple time periods such as quarters or years.
We can determine the portfolio value versus periods, given by
\[
V_t = V_1 \prod_{\tau=1}^{t-1} (1+R^\text{net}_\tau),
\]
where $V_1$ is the portfolio value at period $t=1$. From this we can evaluate
quantities like the average or maximum value of drawdown over quarters or years.

\paragraph{Variations.}
The idea of back-testing or simulating portfolio performance 
can be used for several other tasks.
In one variation on a back-test called a \emph{stress test}, we use 
historical data modified to be more challenging, \eg, lower returns 
or higher costs than actually occurred.

Another variation called \emph{performance forecasting} 
uses data that are simulated or generated, 
starting from the current portfolio
out to some horizon like one year in the future, or the end of 
current fiscal year.
We generate some number of possible future values of quantities such 
as returns, along with the corresponding forecasts of them,
and simulate the performance for each of these.  This gives us an idea of
what we can expect our future performance to be, for example as 
a range of values or quantiles.

Yet another variation is a \emph{retrospective what-if} simulation.
Here we take a live portfolio and go back, say, three months. Starting
from the portfolio holdings at that time,
we simulate forward to the present, after making 
some changes to our trading method, \eg, modifying some parameters.
The fact that the current portfolio value would be higher (according to
our simulation) if the PM had
reduced the target risk three months ago is of course not
directly actionable.  But it still very useful information for the PM.

\paragraph{Parameter tuning.}
Perhaps the most important use of back-testing is to help the PM choose 
parameter values in the Markowitz portfolio construction problem.
While some parameters, like the target risk, are given, others
are less obvious.  
For example, how should we choose $\gamma^\text{trade}$?
The default value of one is our best guess of what the single period 
transaction cost will be.  But perhaps we get better performance
with $\gamma^\text{trade}=2$, which means, roughly speaking,
that we are exaggerating trading cost by a factor of two.
The result, of course, is a reduction in trading compared to
the default value one.  This will result in smaller realized 
transaction costs, but also, possibly, higher return, or smaller
drawdown.  The back-test will reveal what would happen in this case
(to the limits of the back-test accuracy).

To choose among a set of choices for parameters, we carry out
a back-test with each set, and evaluate multiple metrics,
such as realized returns, volatility, and turnover.
Our optimization problem contains target values for these, based
on our forecasts and models; in a back-test we obtain the ex-post
or realized values of these metrics.

To make a final choice of parameters, we must scalarize our
metrics, \ie, create one scalar metric from them,
so we can choose among different sets of parameter values.
For example we might choose to maximize Sharpe ratio,
subject to other metrics being within specified bounds.
Or we could form some kind of weighted combination of the 
individual metrics.

At the very minimum, a PM should always carry out back-tests in
which all of her chosen parameters are, one by one, increased or 
decreased by, say, 20\%.
Even with 10 parameters, this requires only 20 back-tests.
If any of these back-tests results in substantially improved
performance, she will need to explain or defend her choices.

This simple method of changing one parameter at a time can be extended
to carry out a crude but often effective parameter search.  
We cycle over the parameters, increasing or decreasing each and carrying out a 
back-test.  When we find a new set of parameter values that has better performance
than the current set of values, we take it as our new values.  This continues 
until increasing or decreasing each parameter value does not improve performance.

Another traditional method of parameter tuning is \emph{gridding}.
We choose a small number of candidate values for each parameter, and
then carry out a back-test for each combination, evaluating 
multiple performance metrics.  Of course this is 
practical only when we are choosing just a few parameters, and we 
consider only a few candidate values for each one.
Gridding is often carried out with a first crude parameter gridding,
with the candidate values spaced by a factor of ten or so; then, when
good values of these parameters are found, a more refined
grid search is used to focus in on parameters near the good ones
found in the first crude search.
In any case there is no reason to find or specify parameter values 
very accurately; specifying them to even 10\% is not needed.
For one thing, the back-test itself is only an approximation.
To put in a negative light, if a back-test reveals
that $\gamma^\text{trade}=2.1$ works well, but 
that $\gamma^\text{trade}=1.9$ and 
$\gamma^\text{trade}=2.3$ work poorly, it is very unlikely
that our trading method will work well in practice.
Similar to the way we want our trading policy to be robust to variations
in the input data, we also want it to be robust to variation in the
parameters.

More sophisticated parameter search methods can also be used.
Many such methods build a statistical model of the good parameter 
values found so far, and obtain new values to try by sampling from 
the distribution; see, \eg, \cite{maher2022hola} for more discussion.
Another option is to obtain not just the value of some composite metric,
but also its gradient with respect to the parameters.  This very daunting 
computation can be carried out by automatic differentiation
systems that can differentiate through the solution of a convex 
optimization problem, such as CVXPYlayers~\cite{cvxpylayers2019,boyd2020embedded}.

\section{Numerical experiments}\label{s-experiments}
In this section we present numerical experiments that illustrate the ideas and 
methods discussed above.
In the first set of experiments, described in \S\ref{s-taming},
we show the effect of several constraints
and objective terms that serve as effective 
regularizers and improve performance.
In \S\ref{s-tuning} we illustrate how parameter tuning via back-tests can
improve performance, and
in \S\ref{s-scaling} we show how the methods we describe scale with
problem size.

\subsection{Data and back-tests}

\paragraph{Data.}
Throughout the experiments we use the same data set, which is based on the
stocks in the S\&P 100 index. We use daily adjusted close price data from 
2000-01-04 to 2023-09-22. We exclude stocks without data for the entire period,
and acknowledge that this inherent survivorship bias in the data set would make
it unsuitable for a real portfolio construction method, but it is sufficient
for our experiment, which is only concerned with the relative performance of
the different methods.  We end up with a universe of $n=74$ assets.
In addition to the price data, we use bid-ask spread
data to estimate the trading costs, as well as the effective federal funds
rate \cite{FRED} for short term borrowing and lending. 
We make the data set available with the code for reproducibility and experimentation at
\begin{center}
\href{https://github.com/cvxgrp/markowitz-reference}{https://github.com/cvxgrp/markowitz-reference}.
\end{center}

\paragraph{Mean prediction.}
Simple estimates of the means work poorly, so in
the spirit of \cite{BoydKahnMultiPeriod}, we use synthetic return predictions
to simulate a proprietary mean prediction method. For each asset,
the synthetic returns for each day are given by
\[
    \hat{r}_t = \alpha (r_t + \epsilon_t),
\]
where $\epsilon_t$ is a zero-mean Gaussian noise term with variance chosen to
obtain a specified information coefficient and $r_t$ is the mean return of the asset
in the week starting on day $t$.
We take the noise variance to be $\sigma^2(1/\alpha-1)$,
where $\alpha$ is the square of the information coefficient, and $\sigma^2$ is
the variance of the return.
(These mean predictions are done for each asset separately.)
We choose an information coefficient of $\sqrt{\alpha} = 0.15$.
Using this parameterization, the sign of the return is predicted correctly
in 52.1\% of all observations, with this number ranging from 50.3\% to 54.1\%
for the individual assets.

\paragraph{Covariance prediction.}
For the covariance prediction, we use a simple EWMA estimator, \ie, the 
covariance matrix at time $t$ is estimated as
\[
\sigmahat_t = \alpha_{t} \sum_{\tau=1}^{t}\beta^{t-\tau} r_\tau r_\tau^T,
\]
where
\[
\alpha_t=\left( \sum_{\tau=1}^{t}\beta^{t-\tau} \right)^{-1} = \frac{1-\beta}{1-\beta^{t}}
\]
is the normalization constant, and
$\beta \in (0,1)$ is the decay factor.
(We use the second moment as the covariance, since the contribution from the 
mean term is negligible.)
We use a half-life of 125 trading days, which corresponds to a decay factor of
$\beta \approx 0.994$. We note that the specific choice of the half-life does not
change the results of the experiments qualitatively.

\paragraph{Spread.}
Our simulations include the transaction cost associated with bid-ask spread.
In simulation we use the realized bid-ask spread; for the Markowitz problems
we use a simple forecast of spread,
the average realized bid-ask spread over the previous five trading days.

\paragraph{Shorting and leverage costs.}
We use the effective federal funds rate as a proxy for interest on cash for both
borrowing and lending. When shorting an asset we add a 5\% annualized spread 
over the effective federal funds rate to approximate the shorting cost in our
simulation. For forecasting, we set $\kappa^{\text{short}}$ to 7.5\% annualized, 
and $\kappa^\text{borrow}$ to the effective federal funds rate.

\paragraph{Back-tests.}
We use a simple back-test to evaluate the performance of the different
methods. We start with a warm-up period of 500 trading days for our estimators
leaving us with 5,686 trading days, or approximately 22 years of data. The first
1,250 trading days (five years) are used to initialize the priority
parameters. This leaves us with 4,436 out-of-sample trading days,
approximately 17 years.
Starting with an initial cash allocation of \$1,000,000, we call the
portfolio construction method each day to obtain the target weights.
We then execute the trades at the closing price, rebalancing the portfolio to the
new target weights, taking into account the weight changes due to the returns 
from the previous day. 
Buy and sell orders are executed at the ask and bid prices, respectively, and 
interest is paid on borrowed cash and short stocks, and received on cash holdings.

\subsection{Taming Markowitz}\label{s-taming}
In this first experiment we show how a basic Markowitz
portfolio construction method can lead to the undesirable behavior that 
would prompt the alleged deficiencies described in~\S\ref{s-alleged-deficiencies}.
We then show how adding just one more reasonable constraint or objective term
improves the performance, taming the basic Markowitz method.

\paragraph{Basic Markowitz.}
We start by solving the basic Markowitz problem~\eqref{e-markowitz-original}
for each day in the data set, with the risk target set to 10\% annualized volatility.
Unsurprisingly the basic Markowitz problem results in
poor performance, as seen in the second line of table~\ref{t-taming}.
It has low mean return, high volatility (well above
the target 10\%), a low Sharpe ratio,
high leverage and turnover, and a maximum drawdown of almost 80\%.
This basic Markowitz portfolio performs considerably worse than an 
equal-weighted portfolio, which we give as a baseline on the top line of 
table~\ref{t-taming}.

\begin{table}
    \begin{minipage}{\textwidth}
        \centering
        \begin{tabular}{lrrrrrr}
        \hline
        & Return & Volatility & Sharpe & Turnover & Leverage & Drawdown \\ \hline
        Equal weight & 14.1\% & 20.1\% & 0.66 & 1.2 & 1.0 & 50.5\% \\
        Basic Markowitz & 3.7\% & 14.5\% & 0.19 & 1145.2 & 9.3 & 78.9\% \\ \hline
        Weight-limited & 20.2\% & 11.5\% & 1.69 & 638.4 & 5.1 & 30.0\% \\
        Leverage-limited & 22.9\% & 11.9\% & 1.86 & 383.6 & 1.6 & 14.9\% \\
        Turnover-limited & 19.0\% & 11.8\% & 1.54 & 26.1 & 6.5 & 25.0\% \\
        Robust & 15.7\% & 9.0\% & 1.64 & 458.8 & 3.2 & 24.7\% \\ \hline
        Markowitz++  & 38.6\% & 8.7\% & 4.32 & 28.0 & 1.8 & 7.0\% \\
        Tuned Markowitz++ & 41.8\% & 8.8\% & 4.65 & 38.6 & 1.6 & 6.4\% \\
        \hline
        \end{tabular}
        \caption{Back-test results for different trading policies.}
        \label{t-taming}
    \end{minipage}
\end{table}

\paragraph{Markowitz with regularization.}
In a series of four experiments we show how adding just one more reasonable 
constraint or objective term to the basic Markowitz method can greatly improve 
the performance.

In the first experiment we add portfolio weight limits of 10\% for
long positions and -5\% for short positions.  We limit the cash weight
to lie between $-5\%$ and $100\%$ (which guarantees feasibility).
Adding these asset and cash weight limits leads to a significant improvement in the
performance of the portfolio shown in the third row of table~\ref{t-taming},
with the Sharpe ratio increasing to 1.69 (from 0.19),
and the maximum drawdown decreasing to 30\%.  In addition the realized volatility,
11.5\%, is closer to the target value 10\% than the basic Markowitz trading policy.
The turnover is still very high, however,
and the maximum leverage is still large at above 5.

In the second experiment we add a leverage limit to the basic Markowitz problem,
with $L^\text{tar}=1.6$. This one additional constraint also greatly improves
performance, as seen in the fourth row of table~\ref{t-taming}, 
but with a lower
turnover and (not surprisingly) a lower maximum leverage, which is at our
target value $1.6$.

Our third experiment adds a turnover limit of $T^\text{tar}=25$ to the basic 
Markowitz problem.
This additional constraint drops the turnover considerably, to a value
near the target, but still achieves high return, Sharpe ratio, and even lower 
maximum drawdown.

Our fourth experiment adds robustness to the return and risk forecasts.
As simple choices we set all entries of $\rho$ to the 20th percentile of the 
absolute value of the return forecast at each time step, and use $\varrho = 0.02$. 
This robustification also improves performance. Not surprisingly the 
realized risk comes in under our target, since we use the robust risk ex-ante;
we could achieve realized risk closer to our desired target 10\% by increasing the 
target to something like 11.5\% (which we didn't do).

\subsection{Markowitz++}
In the four experiments described above, we see that 
adding just one reasonable additional constraint 
or objective term to the basic Markowitz problem greatly improves the performance.
In our last experiment, we include all of these constraints and terms, with parameters
\[
\begin{array}{l}
\gamma^\text{hold} = 1, \quad
\gamma^\text{trade} = 1, \quad
\sigma^{\text{tar}} = 0.10 \\[0.5em]
c^\text{min} = -0.05, \quad 
c^\text{max} =1.00, \quad 
w^\text{min} = -0.05, \quad 
w^\text{max} =0.10, \quad 
L^\text{tar} =1.6\\[0.5em]
z^\text{min} = -0.10, \quad
z^\text{max} =0.10, \quad 
T^\text{tar}=25.
\end{array}
\]
The mean uncertainty parameter $\rho$ is chosen as the 20th percentile of the 
absolute value of the return forecast, and $\varrho = 0.02$.
We soften the risk target, leverage limit, and turnover limit, using the
priority parameters
\[
\gamma^\text{risk}=5\times 10^{-2}, \quad
\gamma^\text{lev}=5\times 10^{-4}, \quad      
\gamma^\text{turn}=2.5\times 10^{-3}.
\]
These were chosen as the 70th percentiles for the corresponding Lagrange
multipliers of the hard constraints in the basic Markowitz problem for the risk
and turnover limits, and as 25\% of the maximum Lagrange multiplier for the
leverage limit, over the five years leading up to the out-of-sample study. 
(We selected $\gamma^{\text{lev}}$ this way since the
corresponding constraint was active very rarely in the basic Markowitz problem.)

With this Markowitz++ method, we obtain
the performance listed in the second from bottom row of table~\ref{t-taming}.
It is considerably better than the performance achieved by adding just one
additional constraint, as in the four previous experiments, and very much
better than the basic Markowitz method.
Not surprisingly it achieves good performance on all metrics, with a high Sharpe
ratio, reasonable tracking of our volatility target, modest turnover and leverage,
and very small maximum drawdown.
When the parameters are tuned annually, as detailed in the next section,
we see even more improvement, as shown in the bottom row of table~\ref{t-taming}.

The Sharpe ratios on the bottom two rows are high.  We remind the reader that our data has 
survivorship bias and uses synthetic (but realistic) mean return forecasts, so
the performance should not be thought of
as implementable.  But the differences in performance of the 
different trading methods is significant.

\subsection{Parameter tuning}\label{s-tuning}
In this section we show how parameter tuning can be used to improve the 
performance of the portfolio construction method. 
We will tune the parameters
$\gamma^\text{hold}$, 
$\gamma^\text{trade}$,
$\gamma^\text{lev}$,
$\gamma^\text{risk}$,
and $\gamma^\text{turn}$, keeping the other parameters fixed.
We start from the values used in Markowitz++.

\paragraph{Experimental setup.} 
We tune the parameters at the start of every year, on the previous two years of 
data, and then fix the tuned parameters for the following year. To tune the
parameters we use the simple cyclic tuning method described in
\S\ref{s-backtest}. We cycle through the parameters one by one. Each time
a parameter is encountered in the loop, we increase it by 25\%; if this yields
an improvement in the performance (defined below), we keep the new value and continue with the
next parameter; if not, we decrease the parameter by 20\% and check if this
yields an improvement. We continue this process until a full loop through all
parameters does not yield any improvement. By improvement in performance we
mean that all the following are satisfied: 
\BIT 
\item The in-sample Sharpe ratio increases.
\item The in-sample annualized turnover is no more than 50.
\item The in-sample maximum
leverage is no more than 2.
\item The in-sample annualized volatility is no more than 15\%.
\EIT

\paragraph{Results.} 
Tuning the parameters every year
yields the performance given in the last row of table~\ref{t-taming}. 
We see a modest but significant boost in performance over untuned Markowitz++.

\begin{figure}
    \centering
    \begin{subfigure}[b]{0.45\textwidth}
        \includegraphics[width=\linewidth]{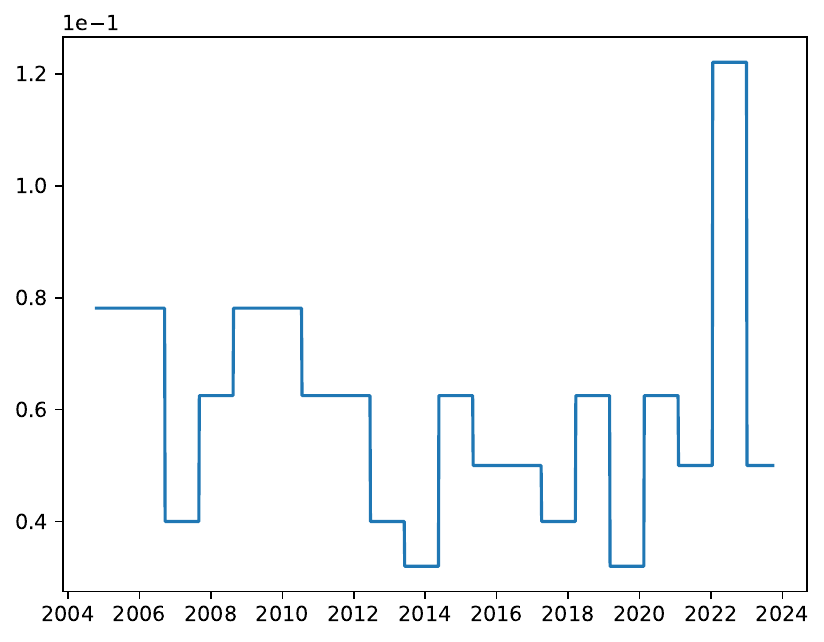}
        \caption{$\gamma^{\text{risk}}$}
    \end{subfigure}
    \begin{subfigure}[b]{0.45\textwidth}
        \includegraphics[width=\linewidth]{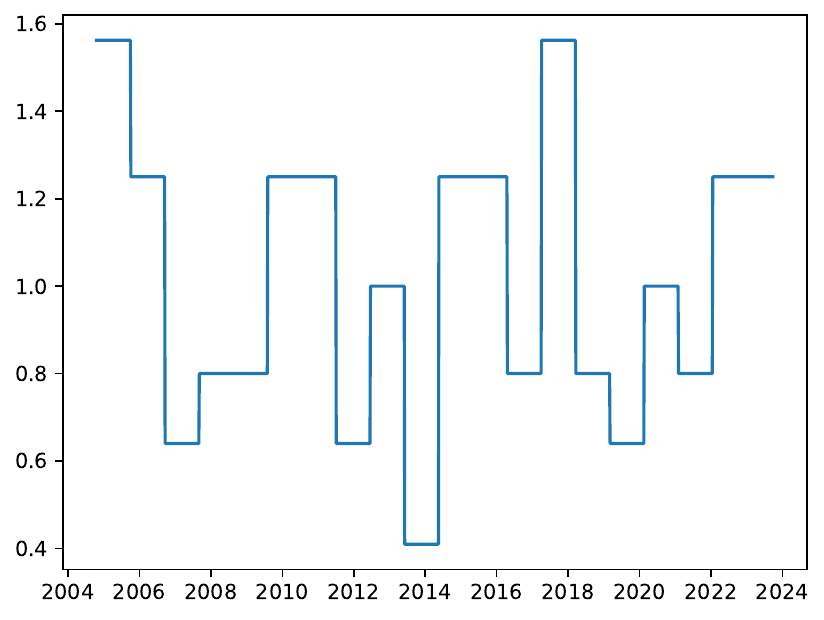}
        \caption{$\gamma^{\text{hold}}$}
    \end{subfigure} \\
    \begin{subfigure}[b]{0.45\textwidth}
        \includegraphics[width=\linewidth]{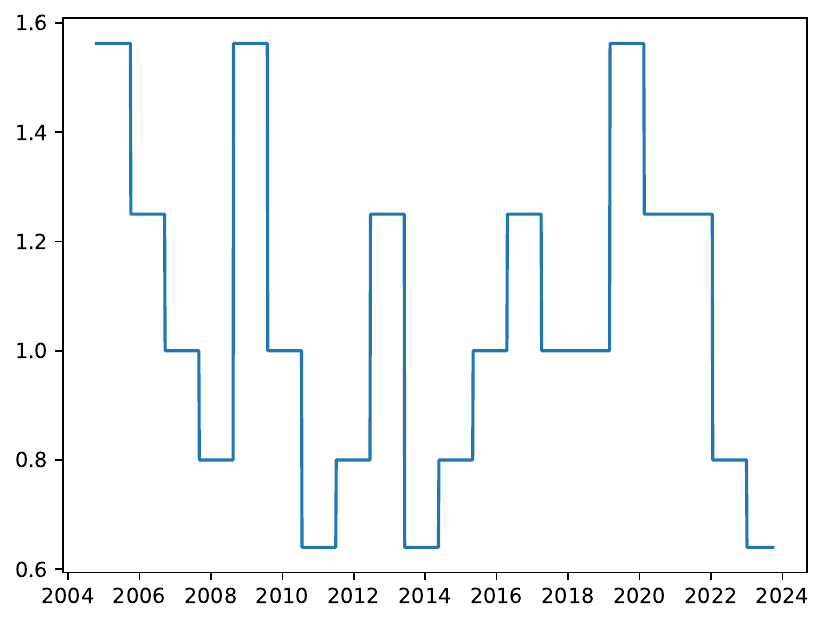}
        \caption{$\gamma^{\text{trade}}$}
    \end{subfigure}
    \begin{subfigure}[b]{0.45\textwidth}
        \includegraphics[width=\linewidth]{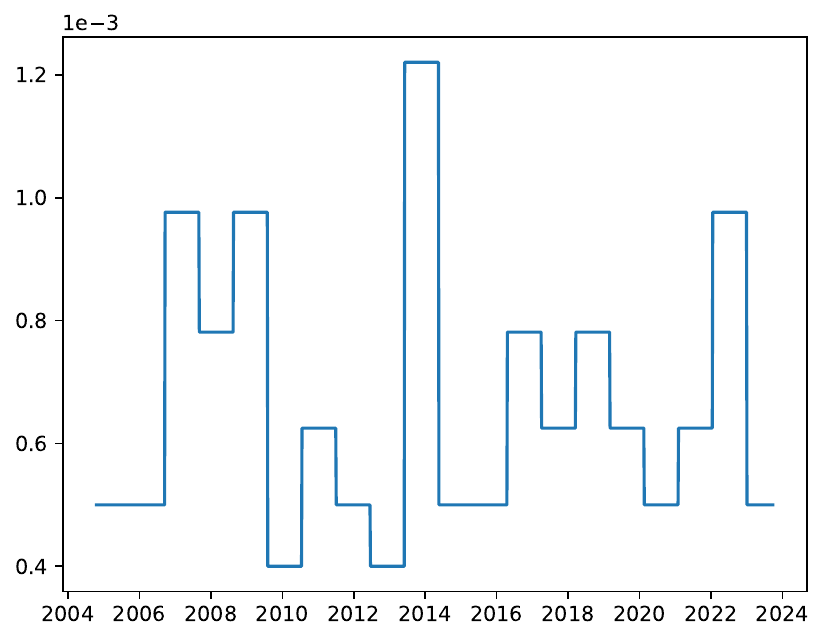}
        \caption{$\gamma^{\text{lev}}$}
    \end{subfigure}\\
    \begin{subfigure}[b]{0.45\textwidth}
        \includegraphics[width=\linewidth]{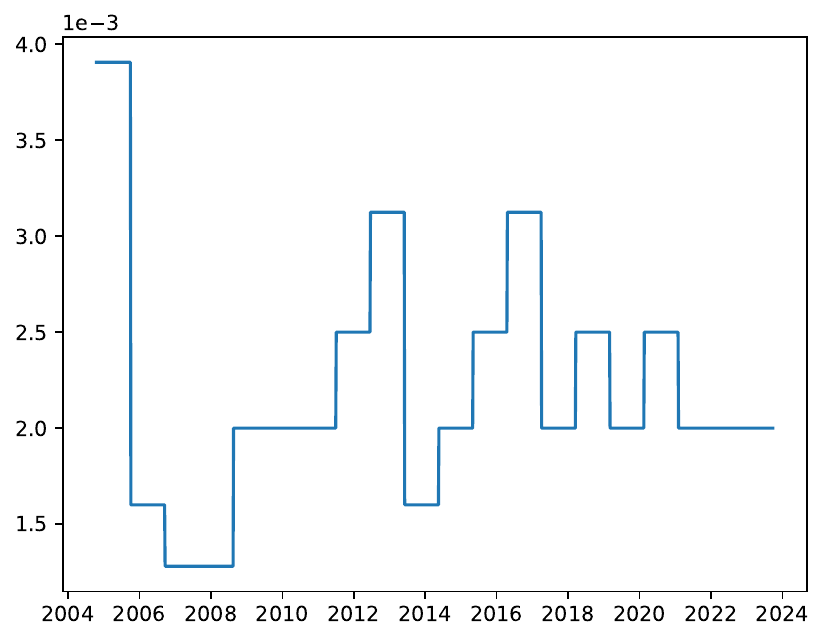}
        \caption{$\gamma^{\text{turn}}$}
    \end{subfigure}
    \caption{Tuned parameters over time.}
    \label{fig:tuning_over_time}
\end{figure}

The tuned
parameters over time are shown in figure~\ref{fig:tuning_over_time}. We can note
several intuitive patterns in the parameter values. For example,
$\gamma^{\text{risk}}$ increases during 2008 to account for the high 
uncertainty in the market during this period. Similarly, $\gamma^{\text{turn}}$ 
decreases during the same period, likely to allow us to trade more freely to
satisfy the other constraints; interestingly $\gamma^{\text{trade}}$ increases
during the same period, likely to push us toward more liquid stocks when trading
increases.
During the same period $\gamma^{\text{lev}}$ increases to reduce
leverage. Similar patterns can be observed in 2020.

\begin{figure}
    \centering
    \begin{subfigure}[b]{0.45\textwidth}
        \includegraphics[width=\linewidth]{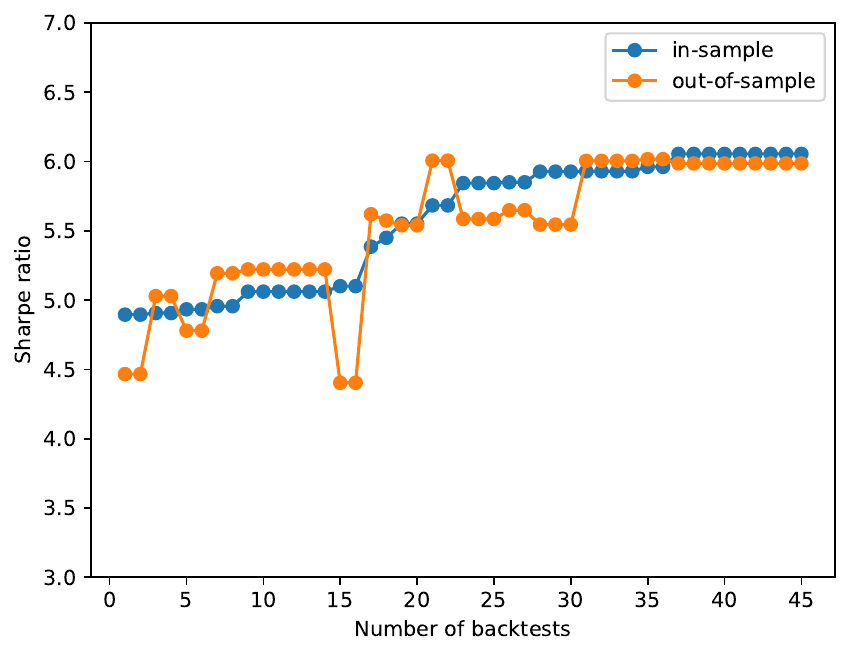}
        \caption{Sharpe ratio.}
    \end{subfigure}
    \begin{subfigure}[b]{0.45\textwidth}
        \includegraphics[width=\linewidth]{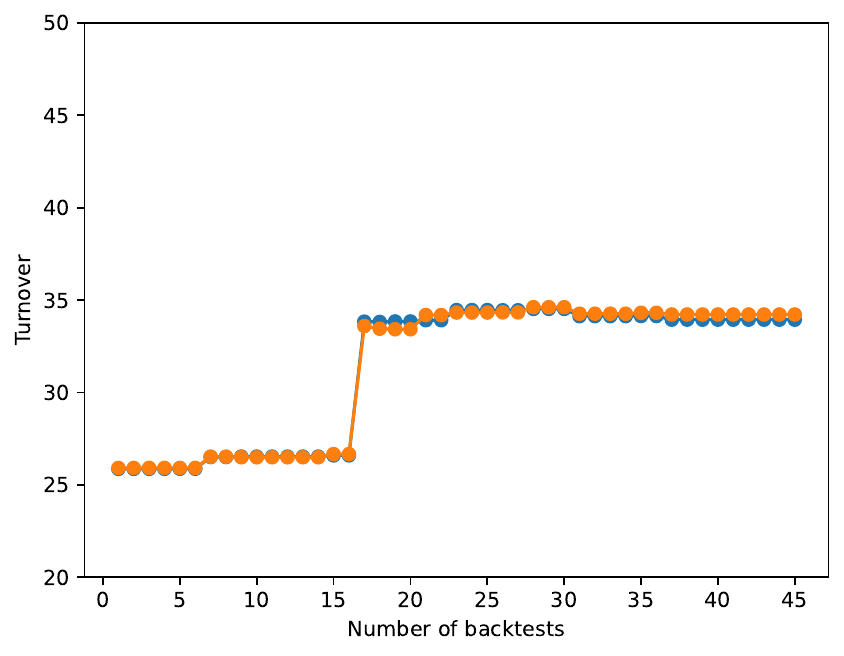}
        \caption{Turnover.}
    \end{subfigure}\\
    \begin{subfigure}[b]{0.45\textwidth}
        \includegraphics[width=\linewidth]{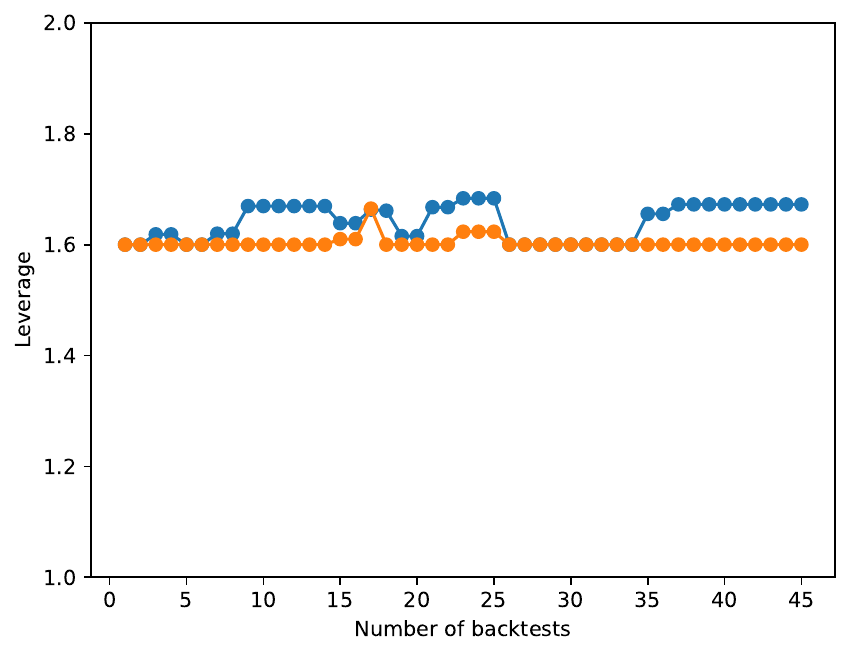}
        \caption{Leverage.}
    \end{subfigure}
    \begin{subfigure}[b]{0.45\textwidth}
        \includegraphics[width=\linewidth]{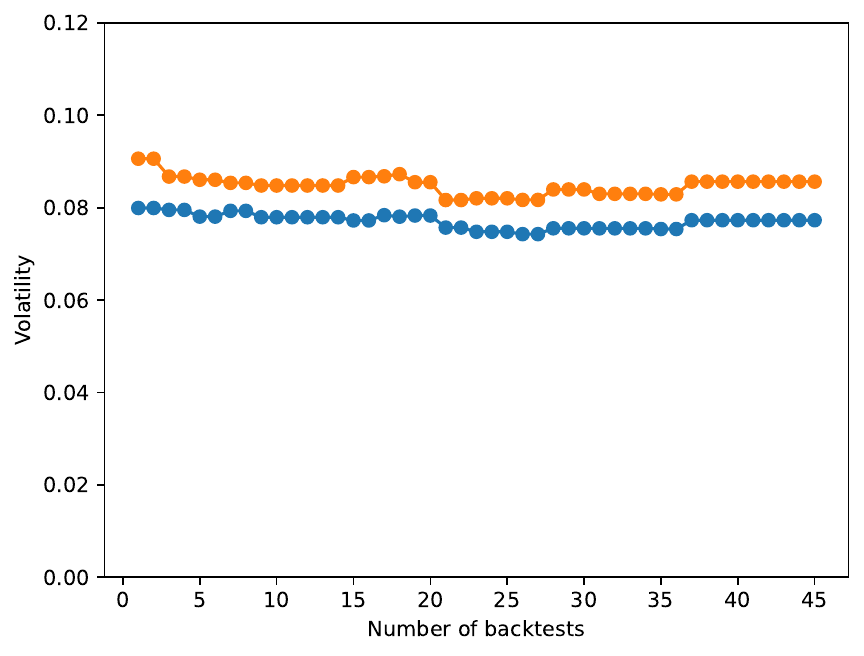}
        \caption{Volatility.}
    \end{subfigure}
    \caption{Parameter tuning results.}
    \label{fig:tuning}
\end{figure}

\paragraph{Tuning evolution.}
Here we show an example of the evolution of tuning, showing both 
in- and out-of-sample values of Sharpe ratio, turnover, leverage, and volatility. 
The in-sample period is April 19, 2016 to March 19, 2018, and the out-of-sample period
March 20, 2018 to March 4, 2019.  These are shown in figure~\ref{fig:tuning}.
This tuning process converged after 45 back-tests to the parameter values
\[
\gamma^\text{risk}=4\times 10^{-2}, \quad
\gamma^\text{hold}=0.64, \quad
\gamma^\text{trade}=0.64, \quad
\gamma^\text{lev}=5\times 10^{-4}, \quad
\gamma^\text{turn}=1.6\times 10^{-3}.
\]
We can see that tuning increases the Sharpe ratio both in- and out-of-sample,
while keeping the leverage, turnover, and volatility reasonable.
In this example we end up changing $4$ of our $5$ adjustable parameters,
although not by much, which
shows that our initial default parameter values were already quite good.
Still, we obtain a significant boost in performance.

\subsection{Annual performance}
\begin{figure}
\centering

\begin{subfigure}{0.45\textwidth}
\centering
\includegraphics[width=\linewidth]{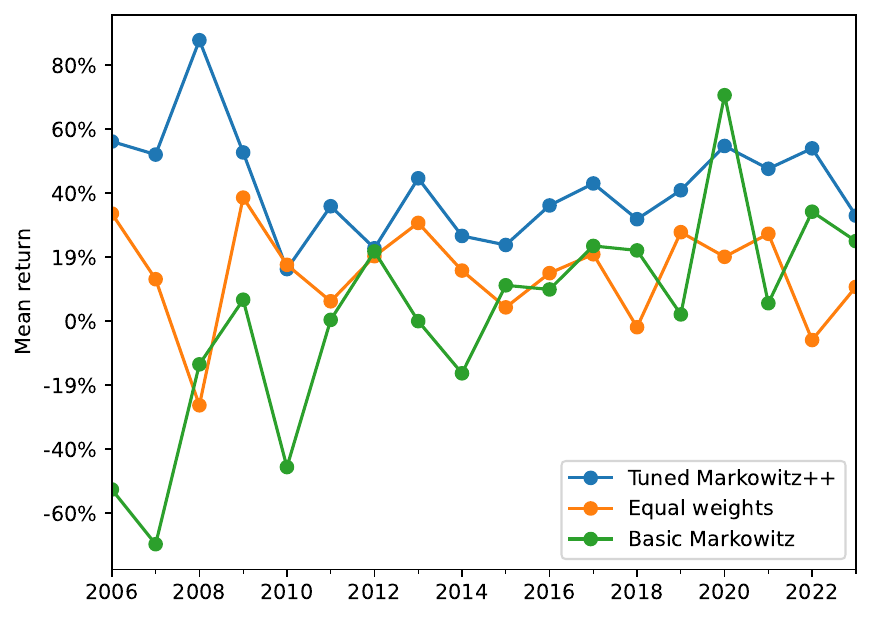}
\caption{Yearly annualized returns.}
\end{subfigure}

\begin{subfigure}{0.45\textwidth}
\centering
\includegraphics[width=\linewidth]{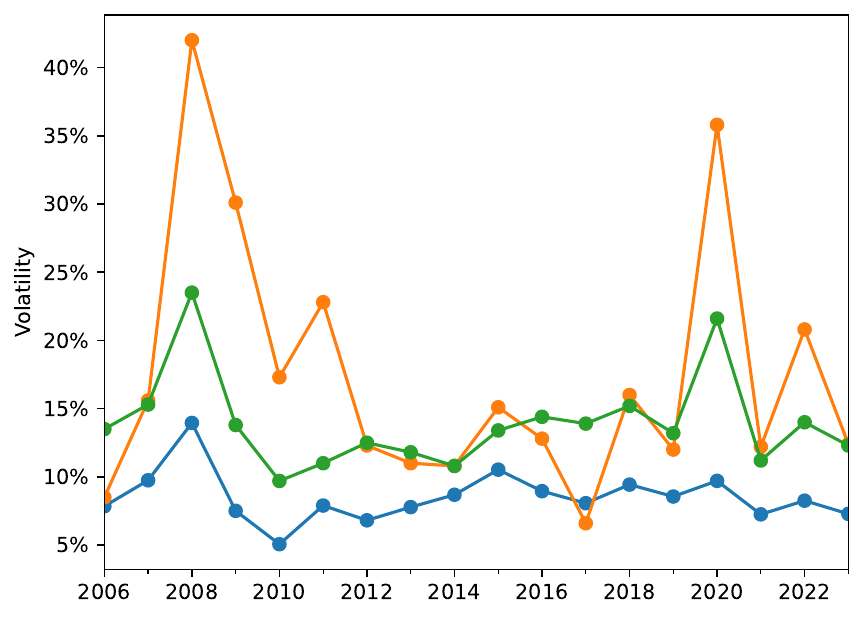}
\caption{Yearly annualized volatilities.}
\end{subfigure}

\begin{subfigure}{0.45\textwidth}
\centering
\includegraphics[width=\linewidth]{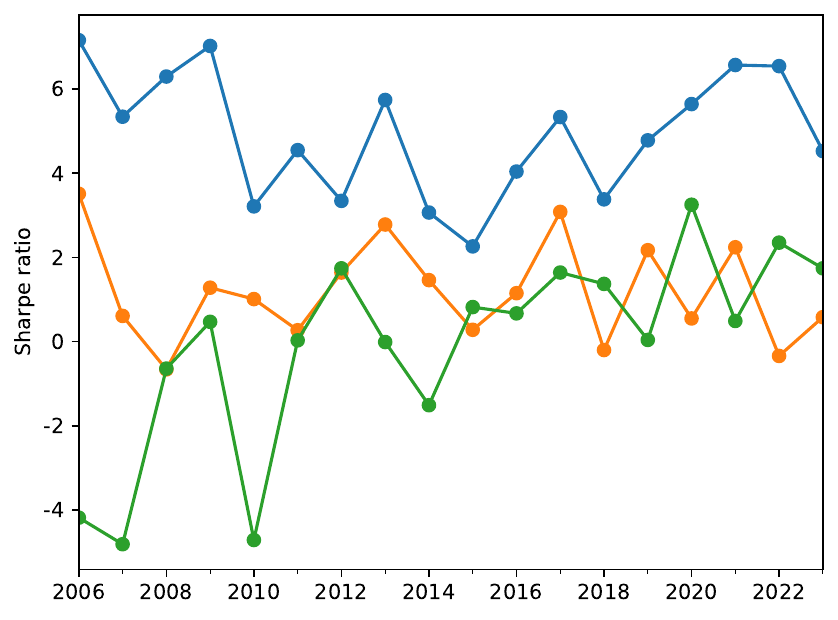}
\caption{Yearly annualized mean Sharpe ratios.}
\end{subfigure}

\caption{Yearly annualized metrics for the equal weight portfolio,
basic Markowitz, and tuned Markowitz++.}
\label{f-annual-performance}
\end{figure}

The performance analyses described above and summarized in table~\ref{t-taming}
give aggregate metrics over a 17 year out-of-sample period, 
long enough to include multiple distinct
market regimes as a well as a few market crashes.
For such a long back-test, it is interesting to see how
the performance in individual years varies with different market regimes.
The realized annual return, volatility, and Sharpe ratio are shown in 
figure~\ref{f-annual-performance}, for basic Markowitz, equal weights, and tuned
Markowitz++.  Here we see that Markowitz++ not only gives 
the performance improvements seen
in table~\ref{t-taming}, but in addition has less variability in performance
across different market regimes.

\clearpage
\subsection{Scaling}\label{s-scaling}
We now turn from the performance of the portfolios to the algorithmic
performance of the portfolio construction method itself.

\begin{figure}
    \centering
    \includegraphics[width=0.8\textwidth]{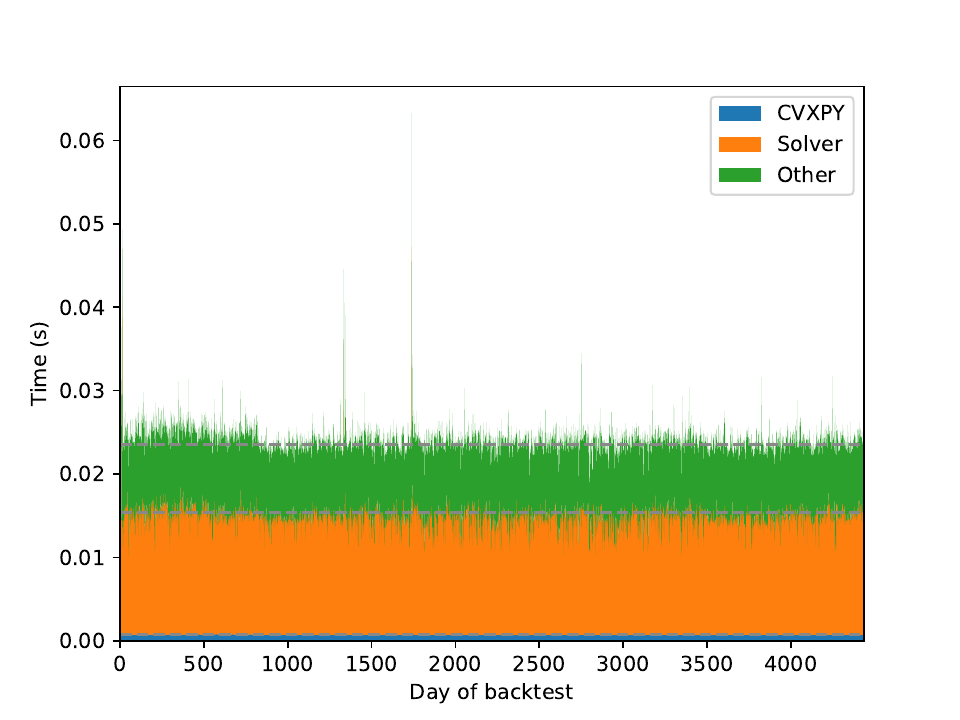}
    \caption{Timing results for the Markowitz problem with a single backtest
    setting.}
    \label{f-timing-parametrized}
\end{figure}

\paragraph{Small problems.}
We start with the small problem used in the previous section, with $n=74$
assets, and without a factor risk model.
Figure~\ref{f-timing-parametrized} shows the time required for 
each of the 4,436 days in the back-test, broken down into
updating and logging (shown in green), 
CVXPY overhead (shown in blue, negligible),
and solver time, the time required to solve the resulting cone program.
(We do not count factorizing the covariance matrix, or computing the mean
forecasts, since these are done ahead of time, and the time is amortized across 
all back-tests.)

The 17 year back-test, which involve solving 4,436 problems, takes
around 104 seconds on a MacBook Pro with an M1 Pro processor, or about 23ms
per day on average.
About 63\% of the time is spent in the
solver, which in this case is MOSEK \cite{aps2020mosek}, with other solvers giving similar
results, including open-source solvers such as ECOS~\cite{domahidi2013ecos}, 
Clarabel~\cite{clarabeldocs}, and SCS~\cite{o2016conic}.
Only 3\% of the time is spent in the compilation step using CVXPY.
The averages for each component of the timings are indicated by the horizontal
lines in the figure.

For a small problem like this one, we can carry out a one-year back-test 
(around 250 trading days) in around six seconds, on a single thread.
A single processor with 32 threads can carry out around 20,000 one-year back-tests 
in an hour.
There is little excuse for a PM who does not carry out many back-tests,
even if only to vary the parameters around their chosen values.

\paragraph{Large problems.}
We now investigate the scaling of the method with problem size. As outlined
in \S\ref{s-factor-model}, a factor model improves the scaling
from $O(n^3)$ to $O(nk^2)$. To illustrate this,
we solve the Markowitz problem for different values of $n$ and $k$ using 
randomly generated but realistic data. Table~\ref{t-scaling} shows the average solve
time for each problem size across 30 instances using MOSEK.
(Solve times with open-source solvers such as Clarabel were a bit longer.)
We can see that even very large problems can be solved with stunning speed.

We solved many more problems
than those shown in table \ref{t-scaling}, and used the solve times to
fit a log-log model, approximating the solve time as $an^b k^c$, with 
parameters $a,b,c$.  We obtained coefficients
$b=0.79$ and $c=1.72$, consistent with the theoretical scaling of $O(nk^2)$.

When the problems are even larger, generic software 
reaches its limits.  In such cases, users may consider switching to first-order
methods like the Alternating Direction Method of Multipliers 
(ADMM)~\cite{moehle2022portfolio, fougnerPOGS, Fougner2018, Parikh2013, boydADMM},
which can offer better scalability and efficiency for very large problems.

\begin{table}
    \begin{minipage}{\textwidth}
        \centering
        \begin{tabular}{rrr}
        \hline
        Assets $n$ & Factors $k$ & Solve time (s) \\ \hline
        100 & 10 & 0.01\\
        500 & 20 & 0.07\\
        500 & 50 & 0.10\\
        2,000 & 50 & 0.23\\
        2,000 & 100 & 0.22\\
        10,000 & 50 & 0.65\\
        10,000 & 100 & 0.89\\
        50,000 & 200 & 9.00\\
        50,000 & 500 & 17.77\\
        \hline
        \end{tabular}
        \caption{Average solve times for Markowitz++ problem for MOSEK,
for different problem sizes.} \label{t-scaling}
    \end{minipage}
\end{table}

\clearpage
\section{Conclusions}
It was Markowitz's great insight to formulate the choice of an investment
portfolio as an optimization problem that trades off multiple objectives,
originally just expected return and risk, taken to be the standard deviation 
of the portfolio return.
His original proposal yielded an optimization problem with an analytical
solution for the long-short case, and a QP for the long-only case,
both of which were tractable to solve (for very small problems) even in the 1950s.
Since then, stunning advances in computer power, together with advances in 
optimization, now allow us to formulate and solve much more complex
optimization problems, that directly handle various practical constraints
and mitigate the effects of forecasting errors.  We can solve these problems
fast enough that very large numbers of back-tests can be carried out, to
give us a good idea of the performance we can expect, and to help choose 
good values of the parameters.
It is hardly surprising that these methods are widely used in quantitative
trading today.

While we have vastly more powerful computers, far better software, 
and easier access to data, than Markowitz did in 1952,
we feel that the more complex Markowitz++ optimization problem 
simply realizes his original idea of an optimization-based
portfolio construction method that takes multiple objectives into account.

\subsection*{Acknowledgments}
The authors thank Trevor Hastie, Mykel Kochenderfer, Mark Mueller, and Rishi Narang
for helpful feedback on an earlier version of this paper. We gratefully
acknowledge support from the Office of Naval Research. This work was partially
supported by ACCESS -- AI Chip Center for Emerging Smart Systems.
Kasper Johansson was partially funded by the Sweden America Foundation.

\clearpage
{\small
\bibliography{refs}
}

\clearpage
\appendix

\section{Coding tricks}
The problem described in \S\ref{s-markowitz-problem} can essentially 
be typed directly into a DSL such as CVXPY, with very few changes.
In this section we mention a few simple tricks in formulating the problem
(for a DSL) that lead to better performance.

\paragraph{Quadratic forms versus Euclidean norms.}
Traditional portfolio construction optimization formulations use
quadratic forms such as $w^T\Sigma w$.
Modern convex optimization solvers can directly handle the Euclidean norm
without squaring to obtain a quadratic form.
Using norm expressions instead of quadratic forms 
is often more natural, and has better numerical properties.
For example a risk limit, traditionally expressed using a quadratic form as 
\[
w^T \Sigma w \leq (\sigma^\text{tar})^2,
\]
is better expressed using a Euclidean norm as
\[
\| L^T w \|_2 \leq \sigma^\text{tar},
\]
where $L$ is the Cholesky factor of $\Sigma$, \ie, $LL^T=\Sigma$,
with $L$ lower triangular with positive diagonal entries.

\paragraph{Exploiting the factor model.}
To exploit the factor model, it is critical to \emph{never} form the covariance
matrix $\Sigma=F\Sigma^\text{f}F^T + D$.  The first disadvantage of doing this 
is that we have to (needlessly) store 
an $n \times n$ matrix, which can be a challenge when $n$ is on the order
of tens of thousands.  In addition, the solver will be slowed by a dramatic factor
as mentioned in \S\ref{s-factor-model}.

To exploit the factor model, we introduce the data matrix $\tilde F=FL$,
where $L$ is the Cholesky factor of $\Sigma^\text{f}$,
so 
$\tilde F\tilde F^T = F\Sigma^\text{f}F^T$.
The portfolio variance is
\[
\sigma^2 = w^T \tilde F \tilde F^T w + w^TDw = \|\tilde F^Tw\|_2^2 + \|D^{1/2} w\|_2^2,
\]
so the risk can be expressed using Euclidean norms as
\[
\sigma = \left\|
\left(
\| \tilde F^Tw \|_2, \|D^{1/2}w\|_2
\right)
\right\|_2.
\]
In this expression, the outer norm is of a $2$-vector;
the inner lefthand norm is of a $k$-vector, and
the inner righthand norm is of an $n$-vector.
Here we should be careful to express $D$ as a diagonal matrix, or 
to express $D^{1/2}w$ as the elementwise (Hadamard) product of two
vectors.

\clearpage
\section{CVXPY code listing} \label{s-CVXPY}
We provide a reference implementation for the problem described in 
\S\ref{s-cvx-formulation}.
This implementation is not optimized for performance, contains no error checking, and is provided
for illustrative purposes only. For a more performant and robust implementation, 
we refer the reader to the
\href{https://github.com/cvxgrp/cvxmarkowitz}{cvxmarkowitz} package~\cite{cvxmarkowitz}.
Below, we assume that the data and parameters are already defined in corresponding data structures.
The complete code for the reference implementation is available at 
\begin{center}
\href{https://github.com/cvxgrp/markowitz-reference}{https://github.com/cvxgrp/markowitz-reference}.
\end{center}

\begin{lstlisting}[language=mypython]
import cvxpy as cp

w, c = cp.Variable(data.n_assets), cp.Variable()

z = w - data.w_prev
T = cp.norm1(z) / 2
L = cp.norm1(w)

# worst-case (robust) return
factor_return = (data.F @ data.factor_mean).T @ w
idio_return = data.idio_mean @ w
mean_return = factor_return + idio_return + data.risk_free * c
return_uncertainty = param.rho_mean @ cp.abs(w)
return_wc = mean_return - return_uncertainty

# worst-case (robust) risk
factor_risk = cp.norm2((data.F @ data.factor_covariance_chol).T @ w)
idio_risk = cp.norm2(cp.multiply(data.idio_volas, w))
risk = cp.norm2(cp.hstack([factor_risk, idio_risk]))
risk_uncertainty = param.rho_covariance**0.5 * data.volas @ cp.abs(w)
risk_wc = cp.norm2(cp.hstack([risk, risk_uncertainty]))

asset_holding_cost = data.kappa_short @ cp.pos(-w)
cash_holding_cost = data.kappa_borrow * cp.pos(-c)
holding_cost = asset_holding_cost + cash_holding_cost

spread_cost = data.kappa_spread @ cp.abs(z)
impact_cost = data.kappa_impact @ cp.power(cp.abs(z), 3 / 2)
trading_cost = spread_cost + impact_cost

objective = (
    return_wc
    - param.gamma_hold * holding_cost
    - param.gamma_trade * trading_cost
)

constraints = [
    cp.sum(w) + c == 1,
    param.w_min <= w, w <= param.w_max,
    L <= param.L_tar,
    param.c_min <= c, c <= param.c_max,
    param.z_min <= z, z <= param.z_max,
    T <= param.T_tar,
    risk_wc <= param.risk_target,    
]

problem = cp.Problem(cp.Maximize(objective), constraints)
problem.solve()
\end{lstlisting}

We start by importing the CVXPY package in line 1 and define the variables of 
the problem in line 3. The variable \verb|w| is the vector of asset weights, 
and \verb|c| is the cash weight. We then define the trade vector \verb|z|, 
turnover \verb|T|, and leverage \verb|L| in lines 5--7 to simplify the notation 
in the remainder of the code.

In the next block we first define the mean return in lines 10--12, 
taking into account the factor and idiosyncratic returns, as well as the 
risk-free rate. We then define the uncertainty in the mean return in line 13, 
which then reduces the mean return to the worst-case return in line 14.

Similarly, the robust risk is obtained in lines 17--21 by first defining the 
factor and idiosyncratic risk components, which are combined to the 
portfolio risk. The uncertainty in the risk, which depends on the asset 
volatilities, is combined with the portfolio risk to obtain the worst-case risk 
in line 21. The holding cost is defined in lines 23--25, followed by the trading cost in 
lines 27--29.

We form the objective function in lines 31--35 by combining the worst-case 
return with the holding and trading costs, weighted by the corresponding 
parameters. The constraints are collected in lines 37--45, starting with the budget 
constraint, followed by the holding and trading constraints, and ending with 
the risk constraint.

Finally, the problem is defined in line 47, combining the objective and constraints. 
It is solved in line 48 by simply calling the \verb|.solve()| method on the 
problem instance, with a suitable solver being chosen automatically.

In only 48 lines of code we have defined and solved the Markowitz problem with all the
constraints and objectives described in \S\ref{s-cvx-formulation}. This
underlines the power of using a DSL such as CVXPY to specify convex 
optimization problems in a way that closely follows the mathematical 
formulation.

\paragraph{Parameters.}
Using parameters can provide both a convenient way to specify the problem,
as well as a way to reduce the overhead of CVXPY when solving multiple instances.
To obtain this speedup requires some restrictions on the problem formulation.
For a precise definition we refer the reader to~\cite{cvxpylayers2019}.
Here we only mention that we require expressions to additionally 
be linear, or affine, in the parameters.
For example, we can use CVXPY parameters to easily and quickly change the
mean return by writing to the the \verb|.value| attribute of the \verb|mean| and
\verb|risk_free| parameters.

\begin{lstlisting}[language=mypython]
mean = cp.Parameter(n_assets)
risk_free = cp.Parameter()

mean_return = w @ mean + risk_free * c
\end{lstlisting}

In some cases, it is necessary to reformulate the problem to satisfy the 
additional restrictions required to obtain the speedup, \eg, by introducing 
auxiliary variables. For convenience, we provide a parametrized implementation 
of the Markowitz problem in the code repository, where these reformulations 
have already been carried out.

\clearpage

\end{document}